\documentclass[letterpaper,11pt]{article}
\pdfoutput=1
\def\thefootnote{\arabic{footnote}}

\usepackage{graphicx}% Include figure files
\usepackage{epsfig}% Include figure files
\DeclareMathAlphabet   {\mathsc}{OT1}{cmr}{m}{sc}

\def\[{\left [}
\def\]{\right ]}
\def\({\left (}
\def\){\right )}
\newcommand{\lang}{\left\langle}
\newcommand{\rang}{\right\rangle}
\newcommand{\lbr}{\left\{}
\newcommand{\rbr}{\right\}}
\newcommand{\beq}{\begin{equation}}
\newcommand{\eeq}{\end{equation}}
\newcommand{\bea}{\begin{eqnarray}}
\newcommand{\eea}{\end{eqnarray}}
\newcommand{\oline}[1]{\overline{#1}}

\newcommand{\wtd}[1]{\widetilde{#1}}
\newcommand{\GeV}      {~\mathrm{GeV}}
\newcommand{\TeV}      {~\mathrm{TeV}}

\newcommand{\EW}       {\mathsc{ew}}

\newcommand{\UV}       {\mathsc{uv}}

\newcommand{\PL}       {\mathsc{pl}}

\newcommand{\GUT}      {\mathsc{gut}}

\newcommand{\order}{\mathcal{O}}

\newcommand{\gappeq}{\mathrel{\rlap {\raise.5ex\hbox{$>$}}
{\lower.5ex\hbox{$\sim$}}}}
\newcommand{\lappeq}{\mathrel{\rlap{\raise.5ex\hbox{$<$}}
{\lower.5ex\hbox{$\sim$}}}}

\newcommand{\Ochi}{\Omega_{\chi} {\rm h}^2}
\newcommand{\be}[1]{\begin{equation}\label{#1}}
\newcommand{\ee}{\end{equation}}
\newcommand{\lp}{\left(}
\newcommand{\rp}{\right)}
\newcommand{\lb}{\left[}
\newcommand{\rb}{\right]}
\newcommand{\lc}{\left\{}
\newcommand{\rc}{\right\}}
\newcommand{\mzeroeff}{M_0^\mathrm{eff}}

\newcommand{\mpl}{M_\mathrm{Pl}}
\newcommand{\mgrav}{m_\mathrm{3/2}}

\newcommand{\lmess}{\Lambda_\mathrm{mess}}
\newcommand{\am}{\alpha_m}
\newcommand{\ag}{\alpha_g}
\newcommand{\psig}{\Phi_{e^+}^\mathrm{sig}}
\newcommand{\pbknd}{\Phi_{e^+}^\mathrm{bknd}}
\newcommand{\ebknda}{\Phi_{e^-}^\mathrm{prim}}
\newcommand{\ebkndb}{\Phi_{e^-}^\mathrm{sec}}
\newcommand{\fp}{f^{\mathrm{PAM}}}
\newcommand{\LSP}{{\widetilde{N}_1}}
\newcommand{\lnf}[2]{\ln\lp\frac{#1}{#2}\rp}
\newcommand{\tq}{\quad\quad\quad}
\newcommand{\wt}[1]{\widetilde{#1}}

\begin{document}
\begin{center}

\vskip .1in {\large \bf Dark Matter Prospects in Deflected Mirage
Mediation}

\vskip .4in Michael~Holmes and Brent~D.~Nelson \vskip .1in

{\em Department of Physics, Northeastern University, Boston, MA
02115} \vskip .1in
\end{center}

\begin{abstract}
\noindent The recently introduced deflected mirage mediation (DMM)
model is a string-motivated paradigm in which all three of the major
supersymmetry-breaking transmission mechanisms are operative. We
begin a systematic exploration of the parameter space of this rich
model context, paying special attention to the pattern of gaugino
masses which arise. In this work we focus on the dark matter
phenomenology of the DMM model as such signals are the least
influenced by the model-dependent scalar masses. We find that a
large portion of the parameter space in which the three mediation
mechanisms have a similar effective mass scale of 1~TeV or less will
be probed by future direct and indirect detection experiments.
Distinguishing deflected mirage mediation from the mirage model
without gauge mediation will prove difficult without collider input,
though we indicate how gamma ray signals may provide an opportunity
for distinguishing between the two paradigms. 
\end{abstract}

\renewcommand{\thepage}{\arabic{page}}
\setcounter{page}{1}
\def\thefootnote{\arabic{footnote}}
\setcounter{footnote}{0}

%%%%%%%%%%%%%%%%%%%%%%%%%%%%%%%%%%%%%%%%%%%%%%%%%%%%%%%%%%%%%%%%%%%%%%
\noindent\section{Introduction}

If supersymmetry is relevant at the electroweak scale then
forthcoming experimental data should reveal the presence of new
states which will be studied extensively at colliders. If the
lightest supersymmetric particle (LSP) is stable it will provide an
excellent candidate for explaining the presence of non-baryonic dark
matter in the cosmos~\cite{Jungman:1995df}. The phenomenology of
these new states -- both at colliders and in cosmological
observations -- will be determined by their interactions with each
other and with the states of the Standard Model. These are
determined, in turn, by the manner in which supersymmetry breaking
is communicated from a hidden sector to the observable
sector~\cite{Chung:2003fi}.

Over the past ten years most studies of supersymmetric phenomenology
have tended to focus on one of three general paradigms for mediation
of supersymmetry breaking to the observable sector. These are
modulus (or ``gravity'')
mediation~\cite{Nilles:1983ge,Kaplunovsky:1993rd}, gauge
mediation~\cite{Giudice:1998bp} and anomaly
mediation~\cite{Giudice:1998xp,Randall:1998uk}. The latter two are
often motivated by bottom-up concerns such as flavor-changing
neutral currents and CP-violation, but the case of modulus mediation
(a special string-motivated case of general supergravity mediation)
can avoid phenomenological constraints as well, provided the modulus
or moduli in question have family-universal couplings to the
observable sector and the underlying string theory has certain
well-motivated isometries~\cite{Gaillard:2005cw,Choi:2008hn}. These
paradigms are typically studied in isolation, assuming that only one
such mechanism dominates the pattern of soft supersymmetry breaking.
As an example one could consider the hugely influential ``Snowmass
Points and Slopes'' benchmark models designed for studies of
collider phenomenology~\cite{Allanach:2002nj}.

This is a reasonable starting point, but recent work suggests that
there are well-motivated reasons to consider pairs of these
mediation mechanisms -- or perhaps all three mechanisms --
simultaneously. The combination of anomaly mediation and gauge
mediation, known as deflected anomaly mediation, was studied soon
after the anomaly-mediated scenario was first
proposed~\cite{Katz:1999uw,Rattazzi:1999qg}. For the combination of
anomaly mediation and modulus mediation to produce comparable
contributions to soft supersymmetry breaking a hierarchy of mass
scales must be engineered. This was first observed in certain
classes of heterotic orbifold
models~\cite{Gaillard:1999yb,Binetruy:2000md,Gaillard:2007jr} but
received increased attention with the advent of KKLT-type moduli
stabilization in D-brane models~\cite{Kachru:2003aw,Grana:2005jc}. A
particular class of Type~IIB string compactifications with
fluxes~\cite{Choi:2004sx} gave rise to the so-called ``mirage
mediation'' models~\cite{Choi:2005ge,Choi:2005uz,Choi:2007ka} whose
phenomenology has been extensively
studied~\cite{Falkowski:2005ck,Baer:2006id,Baer:2006tb,Choi:2006im,Baer:2007eh,Cho:2007fg,Ellis:2007ac}.

Recently the two combinations described above were combined in a
scheme dubbed ``deflected mirage mediation'' (DMM) in which all
three paradigmatic transmission mechanisms are operative and
comparable~\cite{Everett:2008qy}.\footnote{See also the construction
in~\cite{Endo:2008gi}.} While the DMM framework does not result in
fully generalized superpartner mass possibilities -- particulary for
the all-important gaugino sector, as we will see below -- it does
provide a very rich phenomenology. Like the mirage mediation case
which preceded it, some basic elements of the phenomenology of DMM
have been studied within the contexts of specific string-based
frameworks~\cite{Everett:2008ey,Choi:2008vk,Choi:2009jn}. In this
paper we embark on a program to study the following general
question: if all three supersymmetry-breaking mediation mechanism
are present in roughly equal amounts, how will experimental
observation reveal this fact? By phrasing the question in this
manner we hope to study how certain ``underlying principles'' (such
as the mechanism of supersymmetry-breaking mediation) can be
extracted directly from data without reconstructing the full
low-energy Lagrangian~\cite{Binetruy:2003cy,Altunkaynak:2009tg}.

Here we therefore wish to study primarily the gaugino sector, which
reflects fewer model-dependent properties. Indeed, the DMM scenario
continues to be a two-parameter family of gaugino mass patterns,
with one parameter representing deviations from universality and the
other the overall mass scale. In this paper we will consider the
dark matter signals of the deflected mirage mediation scenario,
which are determined largely by the physics of the gaugino sector
alone. Of course {\em full} model-independence is impossible. Such
important properties as the masses of the supersymmetric Higgs
fields and the Higgsino content of the LSP will depend on soft
supersymmetry-breaking scalar masses which are always
model-dependent. We will consider the truly model-dependent
implications for hadron collider physics in a future
publication~\cite{collider}.

In its simplest manifestation the general mediation scenario would
involve three mass scales, one each for the (bulk) gravity
contribution, superconformal anomaly contribution and the gauge
messenger contribution. This can be re-parameterized by a single
overall scale and two dimensionless ratios. Such was the parameter
set adopted by Everett et al.~\cite{Everett:2008qy}. We will review
the construction of the deflected mirage mediation model in
Section~\ref{sec:DMM}, focusing most strongly on the gaugino sector
where the interplay of supersymmetry-breaking mediation mechanisms
is most transparent. We will study how the addition of gauge
mediation alters the allowed parameter space of the theory.
We then turn to dark matter observations in
Section~\ref{sec:signals}, indicating how the deflection arising
from gauge-charged messengers alters the phenomenology relative to
the un-deflected mirage model. Our survey will include a number of
current and future experiments seeking to detect the presence of
relic neutralinos. We will find that regions of the parameter space
which result in a wino-like or mixed Higgsino/gaugino LSP will be
probed in direct detection experiments at the one ton-year scale as
well as in a variety of indirect detection experiments, provided
that the effective mass scales of the various mediation mechanisms
are in the 1~TeV range or less. These are also the areas in which
the thermal relic abundance of the LSP is no greater than the upper
bound inferred from the WMAP data. Regions of the parameter space
with a heavier spectrum, or with a predominantly bino-like LSP, will
prove much more difficult to observe via cosmological or
astroparticle observations. We perform a basic survey of cosmic ray
signals for the DMM model and find that none of the parameter space
gives rise to an adequate explanation for the recent ATIC data,
though some points can give a reasonable fit to the PAMELA data with
boost factors less than $\order(100)$. However, applying such boosts
to the anti-proton flux as well would make these points in conflict
with a lack of signal in the anti-proton data without modification
of the diffusion model. In Section~\ref{sec:compare} we consider the
extent to which the observations discussed in
Section~\ref{sec:signals} will be helpful in distinguishing the
deflected mirage mediation model from the previously-studied mirage
pattern without gauge-charged messengers. Here we will encounter a
``dark matter inverse problem'' in trying to distinguish between the
two cases. We will demonstrate why this occurs and suggest some
possible avenues for future study to resolve this problem.

\noindent\section{Parameterizing Deflected Mirage Mediation}
\label{sec:DMM} In this section we will review the construction of the soft
Lagrangian in deflected mirage mediation. As this material has been
presented more fully elsewhere~\cite{Everett:2008ey,Choi:2009jn} we
will be brief in our discussion of the underlying string-theoretical
context. We begin in Section~\ref{sec:gaugino} with a basic
derivation of the gaugino mass patterns in deflected mirage
mediation in a model-independent manner. This will allow us to
introduce some important notation and outline the parameterizations
we will be using throughout the rest of the work in
Section~\ref{sec:param}. In Section~\ref{sec:space} we complete the
construction by introducing the soft supersymmetry-breaking scalar
masses and trilinear A-terms and explore the allowed regions in the
parameter space. %from both a bottom-up and tow-down viewpoint.

%%%%%%%%%%%%%%%%%%%%%%%%%%%%%%%%%%%%%%%%%%%%%%%%%%%%%%%%%%%%%%%%%%%
\subsection{Gaugino Masses in Terms of Mass Scales}
\label{sec:gaugino}

In this subsection we present the construction of the gaugino mass
patterns without any reference to its possible origin from
string-theoretic considerations. The treatment here closely
parallels that of~\cite{Altunkaynak:2009tg} in the absence of
gauge-charged messengers. For the sake of theoretical clarity we
will work to one-loop order in the renormalization group equations
in this section. We begin by assuming three contributions to the
soft supersymmetry breaking gaugino masses of the MSSM. Let the
contribution from Planck-suppressed operators be universal in form
with a mass scale given by the quantity $M_0$. The contributions
from the superconformal anomaly will be proportional to some ({\em a
priori} different) mass scale given by $M_g$. We assume that these
contributions arise at some effective high-energy scale $\mu_{\UV}$
at which supersymmetry breaking is transmitted from some hidden
sector to the observable sector. It is common to take this scale to
be the GUT scale at which gauge couplings unify, but in string
constructions one might choose a different (possibly higher scale)
at which the supergravity approximation for the effective Lagrangian
becomes valid. Finally, we will include a contribution from some
gauge-charged messenger sector which is proportional to a third mass
scale $\Lambda_{\rm mess}$. As in the original work
of~\cite{Everett:2008qy,Everett:2008ey} we will assume this
messenger sector comes in complete GUT representations of the
Standard Model gauge group so as to preserve the successful gauge
coupling unification of the MSSM. In particular we assume $N_{m}$
copies of $\mathbf{5}$,$\bar{\mathbf{5}}$ representations under
$SU(5)$, which give rise to contributions to soft supersymmetry
breaking gaugino masses at the energy scale $\mu_{\rm mess} <
\mu_{\UV}$.

The full gaugino mass at the high energy boundary condition scale
$\mu_{\UV}$ is given by the
expression~\cite{Gaillard:1999yb,Bagger:1999rd}
\begin{equation} M_a \(\mu_{\UV}\) = M_0
+ g_a^2\(\mu_{\UV}\) \frac{b'_a}{16\pi^2} m_{3/2} \, ,\label{Mahigh}
\end{equation}
where $a=1,2,3$ labels the Standard Model gauge group factors ${\cal
G}_a$ and $m_{3/2}$ is the value of the gravitino mass, given by
$m_{3/2} = \lang e^{K/2} W \rang$. The quantities $b'_a$ are the
beta-function coefficients for the Standard Model gauge groups. In
our conventions these are given by
\begin{equation}
b_a =  -(3 C_a - \sum_i C_a^i) ,  \label{badef}
\end{equation}
where $C_a$, $C_a^i$ are the quadratic Casimir operators for the
gauge group ${\cal G}_a$, respectively, in the adjoint
representation and in the representation of the matter fields
$\Phi^i$ charged under that group. As the summation in~(\ref{badef})
is over all degrees of freedom in the theory present at the scale
$\mu_{\UV} > \mu_{\rm mess}$, it must include the contribution from
the gauge-charged messenger sector. Therefore we may write
\begin{equation} b'_a = b_a + N_{m} \label{baprm}
\end{equation}
with $b_a$ the beta-function coefficients of the MSSM in the absence
of these messenger fields
\begin{equation} \lbr b_1, b_2, b_3\rbr = \lbr \frac{33}{5}, 1, -3
\rbr \, . \label{baSM} \end{equation}
Note that in the absence of the messenger sector ($N_{m} = 0$) if we
take $\mu_{\UV} = \mu_{\GUT}$ then we would have
\begin{equation} g_1^2\(\mu_{\UV}\) = g_2^2\(\mu_{\UV}\) =
g_3^2 \(\mu_{\UV}\) = g_{\GUT}^2 \simeq \frac{1}{2}\, . \label{gGUT}
\end{equation}
In the presence of the gauge-charged messengers the unification of
gauge couplings continues to occur at a scale $\mu_{\GUT} \simeq 2
\times 10^{16}\GeV$ but the value of the unified gauge coupling
differs from that of~(\ref{gGUT}) by the relation
\begin{equation} \frac{1}{g_a^2\(\mu_{\UV}\)} = \frac{1}{g^2_{\GUT}} -
\frac{N_{m}}{8\pi^2} \ln \(\frac{\mu_{\GUT}}{\mu_{\rm mess}}\)
\label{grelate} \end{equation}
which is valid at one-loop order for all values of $a$.

We now imagine evolving the expressions in~(\ref{Mahigh}) to the
scale $\mu_{\rm mess}$ via the one-loop renormalization group
equations. The first term in~(\ref{Mahigh}) is renormalized by a
multiplicative factor at one loop
\begin{equation} M^{\rm term\, 1}_a \(\mu_{\rm mess}\) = M_0 \[1 -
g_a^2\(\mu_{\rm mess}\) \frac{b'_a}{8\pi^2}
\ln\(\frac{\mu_{\UV}}{\mu_{\rm mess}}\)\] \, , \label{piece1}
\end{equation}
while the second term in~(\ref{Mahigh}) can be evolved by simply
replacing the gauge coupling with its value at the intermediate mass
scale
\begin{equation} M^{\rm term\, 2}_a \(\mu_{\rm mess}\) = g_a^2 \(\mu_{\rm mess}\)
\frac{b'_a}{16\pi^2} m_{3/2} \, . \label{piece2} \end{equation}
Combining~(\ref{piece1}) and~(\ref{piece2}) we have
\begin{equation} M_a \(\mu_{\rm mess}\) = g_a^2 \(\mu_{\rm mess}\)
\frac{b'_a}{16\pi^2} m_{3/2} + M_0 \[1 - g_a^2\(\mu_{\rm mess}\)
\frac{b'_a}{8\pi^2} \ln\(\frac{\mu_{\UV}}{\mu_{\rm mess}}\)\] \, .
\label{Maint} \end{equation}
Clearly, if the two parts of the expression in~(\ref{Maint}) are to
be of roughly the same size it will be necessary to engineer a
situation in which $m_{3/2} \simeq 16\pi^2 M_0$.

At this intermediate scale we may now integrate out the
gauge-charged messenger sector, which in the presence of
supersymmetry breaking parameterized by $m_{3/2}$ generates a
threshold correction to the gaugino masses of the
form~\cite{Pomarol:1999ie}
\begin{equation} \Delta M_a = -N_{m} \frac{g_a^2\(\mu_{\rm
mess}\)}{16\pi^2} \(\Lambda_{\rm mess} + m_{3/2}\)\, .
\label{Mathresh}
\end{equation}
Evolving the combination $M_a \(\mu_{\rm mess}\) + \Delta M_a$ from
the intermediate scale $\mu_{\rm mess}$ to the low-energy scale
$\mu_{\EW}$ (again at one-loop) results in the final expression
\begin{eqnarray}
M_a \(\mu_{\EW}\) &=& g_a^2 \(\mu_{\EW}\) \frac{b_a}{16\pi^2}
m_{3/2} +
\[1-g_a^2 \(\mu_{\EW}\)\frac{b_a}{8\pi^2}\ln\(\frac{\mu_{\rm
mess}}{\mu_{\EW}}\)\] \nonumber \\
& & \times\[M_0 \(1 - g_a^2\(\mu_{\rm mess}\) \frac{b'_a}{8\pi^2}
\ln\(\frac{\mu_{\UV}}{\mu_{\rm mess}}\)\) -N_{m} \frac{
g_a^2\(\mu_{\rm mess}\)}{16\pi^2} \Lambda_{\rm mess}\]\, .
\label{Mafull}
\end{eqnarray}
From~(\ref{Mafull}) we immediately see that we must engineer the
relations $\Lambda_{\rm mess} \simeq m_{3/2} \simeq 16\pi^2 M_0$ if
all three contributions to the gaugino masses are to be comparable
in magnitude. Let us note before going forward that the preceding
derivation made tacit assumptions about the relative phases between
the various contributions in the final expression~(\ref{Mafull}).
Given a true top-down construction these phases should be determined
by the same mechanism which stabilizes the auxiliary fields in the
supergravity Lagrangian.

%%%%%%%%%%%%%%%%%%%%%%%%%%%%%%%%%%%%%%%%%%%%%%%%%%%%%%%%%%%%%%%%%%%%%%%
\subsection{A Theory-Guided Parametrization}
\label{sec:param}

The masses in~(\ref{Mafull}) depend explicitly on the three mass
scales $M_0$, $m_{3/2}$ and $\Lambda_{\rm mess}$ as well as the
number of messenger multiplets $N_{m}$. They also depend implicity
on the value of the intermediate scale $\mu_{\rm mess}$ through the
logarithms and gauge couplings. This somewhat unwieldy expression
can be simplified by relating the explicit mass scales to some
overall mass scale via two dimensionless ratios. Following Everett
et al. we can define
\begin{equation} \alpha_g \equiv \frac{\Lambda_{\rm mess}}{m_{3/2}}
\label{alphag} \end{equation}
which will naturally be of order unity for the cases of interest to
us here. To obtain another ratio of $\order(1)$ it is common to
define
\begin{equation} \alpha_m \equiv \frac{m_{3/2}}{M_0
\ln\(M_{\PL}/m_{3/2}\)} \, , \label{alpham} \end{equation}
where $M_{\PL}$ is the reduced Planck mass $M_{\PL} = 2.4 \times
10^{18} \GeV$. Again, for the models of interest to us here we will
have $\alpha_m$ typically of order unity. Using the
definitions~(\ref{alphag}) and~(\ref{alpham}) the expression
in~(\ref{Mafull}) can be re-expressed as
\begin{eqnarray}
\frac{M_a \(\mu_{\EW}\)}{M_0} &=& g_a^2 \(\mu_{\EW}\)
\alpha_m\frac{b_a}{16\pi^2} \ln\(M_{\PL}/m_{3/2}\) +
\[1-g_a^2 \(\mu_{\EW}\)\frac{b_a}{8\pi^2}\ln\(\frac{\mu_{\rm
mess}}{\mu_{\EW}}\)\] \nonumber \\
& & \times\[1 - g_a^2\(\mu_{\rm mess}\) \frac{b'_a}{8\pi^2}
\ln\(\frac{\mu_{\UV}}{\mu_{\rm mess}}\) - \alpha_m \alpha_g N_{m}
\frac{g_a^2\(\mu_{\rm mess}\)}{16\pi^2} \ln\(M_{\PL}/m_{3/2}\)
\]\, . \label{MaE}
\end{eqnarray}
With these conventions one can define the so-called ``mirage scale''
$\mu_{\rm mir}$ at which the three soft supersymmetry breaking
gaugino masses have an identical value as
\begin{equation} \mu_{\rm mir} = \mu_{\GUT}
\(\frac{m_{3/2}}{M_{\PL}}\)^{\alpha_m \rho/2} \, , \label{mirscale}
\end{equation}
where the parameter $\rho$ is given by
\begin{equation} \rho=\[1+\frac{N_{m}
g_a^2\(\mu_{\UV}\)}{8\pi^2}\lnf{\mu_{\GUT}}{\mu_{\rm
mess}}\]/\[1-\frac{\am\ag
N_{m}g_a^2\(\mu_{\UV}\)}{16\pi^2}\lnf{\mpl}{\mgrav}\]\, .
\label{rho}
\end{equation}

The expression in~(\ref{MaE}) is still somewhat cumbersome. It was
noted by Choi~\cite{Choi:2008vk} that the gaugino mass
expressions~(\ref{Mafull}) at the low scale $\mu_{\EW}$ can be
re-arranged into the form
\begin{equation} M_a(\mu_{\EW})=\mzeroeff \lc 1 + \beta_a(\mu_{\EW})
\ln\lp\frac{\mu_{\EW}}{\mu_{\rm mir}}\rp \rc \, , \label{MaC}
\end{equation}
where we have introduced the variable
\begin{equation} \beta_a(\mu) = \frac{b_a g_a^2(\mu)}{8\pi^2}
\label{beta} \end{equation}
and have defined the effective mass scale
\begin{equation} \mzeroeff = R M_0\,; \quad R = 1 - \frac{N_{m} g_{\GUT}^2}{8\pi^2}
\lc \frac{\alpha_m\alpha_g}{2}\ln\lp\frac{\mpl}{\mgrav}\rp +
\ln\lp\frac{\mu_{\GUT}}{\mu_{\rm mess}}\rp \rc \, . \label{Rdef}
\end{equation}
The expression in~(\ref{MaC}) is precisely identical to that
of~(\ref{MaE}) at one loop order, provided we remember that the
high-scale gauge coupling $g_{\GUT}^2$ appearing in~(\ref{Rdef}) is
related to the high-scale gauge coupling $g_a^2\(\mu_{\UV}\)$
appearing in~(\ref{MaE}) via~(\ref{grelate}). We note that the
dimensionless quantity $R$ is simply the inverse of the parameter
$\rho$ in~(\ref{rho}), and thus
\begin{equation} \mu_{\rm mir} = \mu_{\GUT}
\(\frac{m_{3/2}}{M_{\PL}}\)^{\alpha_m /2R}\, . \end{equation}

In~\cite{Choi:2009jn} the expression in~(\ref{MaC}) was further
modified by introducing two new dimensionless variables based on the
quantity $R$ in~(\ref{Rdef})
\begin{equation} x=\frac{1}{R+\alpha_m}\, , \tq
y=\frac{\alpha_m}{R+\alpha_m} \label{xydef} \end{equation}
which can be used to write~(\ref{MaC}) in the form
\begin{equation} M_a(\mu) =M_0\lp 1+\beta_a(\mu)t \rp x^{-1}\lc 1+y\lb
\frac{\beta_a(\mu)t'}{1+\beta_a(\mu)t} -1 \rb \rc \label{Ma2} \,
,\end{equation}
where we have defined two scaling variables $t$ and $t'$ via
\begin{equation} t=\lnf{\mu}{\mu_{\GUT}}\, , \tq
t'=\frac{1}{2}\lnf{\mpl}{\mgrav}\, . \label{tdef} \end{equation}
The expression in~(\ref{Ma2}) is intended to be evaluated at the
energy scale $\mu = \mu_{\EW} < \mu_{\rm mess}$. The free parameters
in~(\ref{Ma2}) are $\lbr M_0,\,x,\,y\rbr$, though~(\ref{Ma2}) is
still ultimately a function of only one scale and one dimensionless
variable (as we will explicitly show in Section~\ref{sec:compare}).
Having specified $x$ and $y$ values one may obtain $\alpha_m = y/x$,
and then the value of $m_{3/2}$ can be obtained from that of $M_0$
by finding the solution to the definitional equation for $\alpha_m$
in~(\ref{alpham}). Though the series of definitions in~(\ref{Rdef})
and~(\ref{xydef}) may appear to sacrifice clarity for the sake of
brevity, they allow an extremely powerful way to survey all possible
mediation mechanisms in a simple two-dimensional plane.

%%%%%%%%%%%%%%%%%%%%%%%%%%%%%%%%%%%%%%%%%%%%%%%%%%%%%%%%%%%%%%%%%%%%%%
\subsection{Exploring the Parameter Space}
\label{sec:space}

As we have seen, the gaugino mass sector in deflected mirage
mediation is particularly transparent. Though the three soft
supersymmetry-breaking gaugino masses are sufficient to determine a
large part of the dark matter phenomenology of the LSP neutralino,
the scalar sector will also contribute in important, indirect ways.
These include determining the $\mu$-parameter via the electroweak
symmetry-breaking (EWSB) conditions and thereby the LSP
wave-function, and by allowing certain important resonant
annihilation or co-annihilation processes in the early universe.
Therefore we cannot adequately discuss the prospects for dark matter
signals in deflected mirage mediation without specifying the rest of
the soft supersymmetry breaking Lagrangian, particularly the scalar
masses and trilinear couplings.

Doing so requires working within a model framework. We will here
very briefly outline that framework and the relevant expressions
needed to go forward. Further background can be found in the
relevant references, particularly~\cite{Everett:2008ey,Choi:2009jn}.
We imagine Type~IIB string theory compactified on a Calabi-Yau
orientifold. To this is added flux in the manner of KKLT, producing
a region of highly warped geometry. The remainder of the compact
space remains (approximately) a Calabi-Yau manifold. These
background fluxes serve two purposes: the first is to stabilize most
of the geometrical moduli associated with the construction. The
second (and related) purpose is to produce a constant in the
superpotential for the remaining modulus of sufficient size to
produce stabilization with $\alpha_m \sim \order(1)$. This remaining
modulus is a K\"ahler modulus which can be stabilized by
field-theoretic mechanisms such as gaugino condensation.
The observable sector ({\em i.e.} the MSSM) resides on $D_3$ or
$D_7$ branes (or both) in the bulk of the internal manifold. In this
minimal set-up it is plausible that a conspiracy of scales can be
achieved in which direct modulus-mediation (via non-vanishing
$F$-terms for the K\"ahler modulus) and loop-induced
anomaly-mediation are competitive contributions to observable sector
scalar masses.

Let us denote the (chiral) K\"ahler modulus superfield by $T$ and
its highest auxiliary component by $F_T$. If the gauge kinetic
functions $f_a$ for the three Standard Model gauge groups are
universally given by $f_a = T$, then a non-vanishing vacuum value
for $F_T$ will generate a universal component for the gaugino masses
as in~(\ref{Mahigh})
\begin{equation} M_0 = \lang \frac{F_T}{t+\bar{t}} \rang \, ,
\label{FT} \end{equation}
where $t = T|_{\theta =0}$ is the lowest (scalar) component of the
superfield $T$. The quantity in~(\ref{FT}) will also generate scalar
masses and trilinear scalar couplings. The exact form depends on how
the K\"ahler modulus appears in the K\"ahler metric for the matter
superfields, if at all. Such couplings are difficult to compute, but
their leading behavior at large volume (large $\lang t+\bar{t}
\rang$) can be readily inferred and parameterized. We will assume a
K\"ahler potential of the form
\begin{equation} K = \sum_i (T+\oline{T})^{-n_i} \oline{\Phi}_i
\Phi_i\, , \label{K} \end{equation}
with modular weight $n_i$ for the matter superfield $\Phi_i$. These
numbers depend on the location of the observable sector matter
fields and may be vanishing. It was argued in~\cite{Ibanez:2004iv},
using the work of~\cite{Ibanez:1998rf}, that for intersecting
$D$-brane models in the Type~IIB context one should expect $n_i = 0$
if $\Phi_i$ is localized on $D_3$ branes, $n_i = 1$ if $\Phi_i$ is
localized on $D_7$ branes, and $n_i = 1/2$ for matter localized at
the intersection of branes. We point out, however, that no explicit
model realizing both the fluxed stabilization mechanism and the MSSM
matter content has yet been constructed.

In the absence of gauge-charged messengers, the contributions
of~(\ref{FT})
% and~(\ref{FC})
and the assumption in~(\ref{K}) are sufficient to specify the form
of the scalar sector soft supersymmetry breaking terms at a scale
$\mu_{\UV}$. Such a system defines the mirage mediation (or mixed
modulus/anomaly mediation) scenario. To this we now wish to add
$N_m$ copies of gauge-charged messengers in the representation
$\mathbf{5}$,$\bar{\mathbf{5}}$ under $SU(5)$. The presence of such
terms introduces new contributions to the soft scalar masses and
gaugino masses at the scale $\mu_{\rm mess}$ at which these fields
are integrated out of the spectrum. The calculation of such terms
follows the general procedure of gauge mediation
models~\cite{Giudice:1998bp}.
Soft supersymmetry breaking is transmitted to the observable sector
states via loop diagrams involving the messengers. The size of these
contributions is determined by the parameter $\Lambda_{\rm mess}$
appearing in~(\ref{Mathresh}).
In principle this mass scale can be of any size as it is {\em a
priori} independent of the physics of moduli stabilization. The
relative size of $\Lambda_{\rm mess}$ to $m_{3/2}$ (and hence the
parameter $\alpha_g$) will depend on the model one postulates for
dynamically generating masses and splittings for the messenger
sector. Circumstances for which $\alpha_g$ is of order unity are not
hard to construct~\cite{Everett:2008qy,Pomarol:1999ie,Kane:2002qp}.

The soft terms for the scalar masses and trilinear couplings, in the
presence of all three supersymmetry-breaking transmission
mechanisms, are given at the initial scale $\mu_{\UV}
> \mu_{\rm mess}$ by~\cite{Everett:2008ey}
\begin{eqnarray}
A_{ijk}(\mu_{\UV}) & = & M_0 \[(3-n_i -n_j - n_k) -
\frac{(\gamma_i+\gamma_j +
\gamma_k)}{16\pi^2}\alpha_m\ln\(M_{\PL}/m_{3/2}\)\]
\label{AtermUV} \\
m_i^2(\mu_{\UV}) & = & M^2_0\[(1-n_i) -
\frac{\theta'_i}{16\pi^2}\alpha_m\ln\(M_{\PL}/m_{3/2}\) -
\frac{\dot{\gamma'_i}}{(16\pi^2)^2}\(\alpha_m\ln\(M_{\PL}/m_{3/2}\)\)^2
\]\, , \label{scalarUV} \end{eqnarray}
with $\alpha_m$ being the parameter introduced in~(\ref{alpham}) and
the quantities $\gamma_i$, $\theta'_i$ and $\dot{\gamma}'_i$ being
functions of Standard Model gauge and Yukawa couplings. Their
definitions and explicit values can be found
in~\cite{Everett:2008ey}. The primes on the quantities
in~(\ref{scalarUV}) indicate that these parameters are computed
taking into account the presence of the messenger sector as well as
the MSSM matter content. At the scale $\mu = \mu_{\rm mess}$ the
messenger sector is integrated out of the theory, producing the
contribution~(\ref{Mathresh}) for the gaugino masses
\begin{equation}
\Delta M_a = -N_m M_0 \frac{g_a^2(\mu_{\rm mess})}{16\pi^2}
\alpha_m(1+\alpha_g)\ln\(M_{\PL}/m_{3/2}\) \, , \label{DelMa}
\end{equation}
and a contribution
\begin{equation}
\Delta m_i^2 = M_0^2 \sum_a (2N_m C_a) \frac{g_a^4(\mu_{\rm
mess})}{(16\pi^2)^2}\[\alpha_m(1+\alpha_g)\ln\(M_{\PL}/m_{3/2}\)\]^2
\label{DelMi} \end{equation}
for the scalar masses. These expressions can be re-expressed in
terms of the parameter $R$ of~(\ref{Rdef}) and the set $\lbr
x,\,y\rbr$ of~(\ref{xydef}), but doing so is not particularly
illuminating.

Unlike the case of gaugino masses, no analytic expression at low
scales is available for the scalar masses and trilinear couplings --
even at one loop order -- except in certain
approximations~\cite{Choi:2009jn}. Therefore when we wish to
consider the full model-dependent set of soft parameters we must
connect energy scales using the renormalization group (RG)
equations. We do so in two stages. The first connects $\mu_{\UV} =
\mu_{\GUT}$ and $\mu_{\rm mess}$ with boundary conditions given
by~(\ref{Mahigh}), (\ref{AtermUV}) and~(\ref{scalarUV}). At this
scale the corrections~(\ref{DelMa}) and~(\ref{DelMi}) are added and
the soft terms are evolved from $\mu_{\rm mess}$ to $\mu_{\EW} =
M_Z$. From here we pass the low-scale soft parameters to {\tt
SuSpect 2.4}~\cite{Djouadi:2002ze} to check for proper electroweak
symmetry breaking and to compute the physical masses of the various
superpartners.

We will investigate the parameter space defined by the set $\lbr
x,\,y,\,M_0 \rbr$ by considering slices of the $\lbr x,\,y\rbr$
plane for various values of $M_0$. We do this in two different
regimes. In the first case we investigate the explicit model
described in this section for the case where $\tan\beta = 10$,
$\mu_{\rm mess} = 10^{10} \GeV$, $N_m = 3$ and where the modular
weights of the various MSSM states are given by
\begin{equation} \lbr n_Q,\,n_U,\,n_D,\,n_L,\,n_E,\,n_{H_u},\,n_{H_d} \rbr =
\lbr 1/2,\,1/2,\,1/2,\,1/2,\,1/2,\,1,\,1 \rbr \, . \label{weights}
\end{equation}
This choice of modular weights is not motivated by any particular
construction; we choose it to match with certain benchmark models
from the literature. We will call this the ``model-dependent
scenario.'' In the second case we will dispense with the
model-dependent soft-terms by simply setting all trilinear couplings
to zero, set $\mu = m_A = 1 \TeV$ and set all scalar masses to the
value of the gluino soft mass $M_3$ or to 1~TeV, whichever is
larger. We do this by hand at the low energy scale. While
artificial, this provides some sense of how the phenomenology of the
gaugino sector alone is influenced by the choice of $\lbr x,\,y\rbr$
for a given value of $M_0$. We will call this second case the
``model-independent scenario.''

%=(1)============ Allowed Region: Model-Dependent ===============
%\begin{comment}
\begin{figure}[p]
\begin{center}
\includegraphics[scale=0.535]{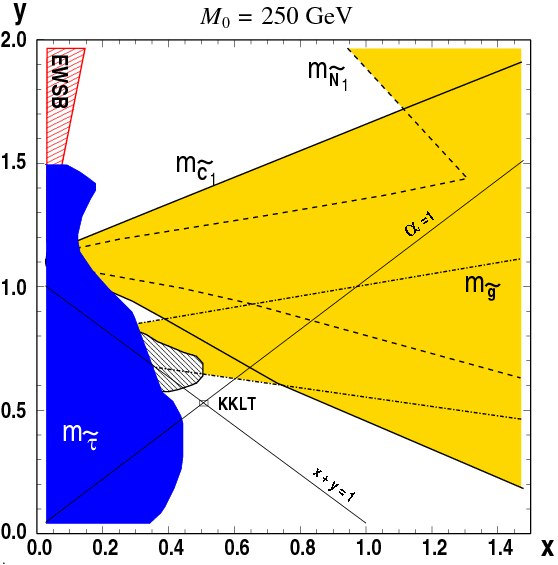}
\includegraphics[scale=0.535]{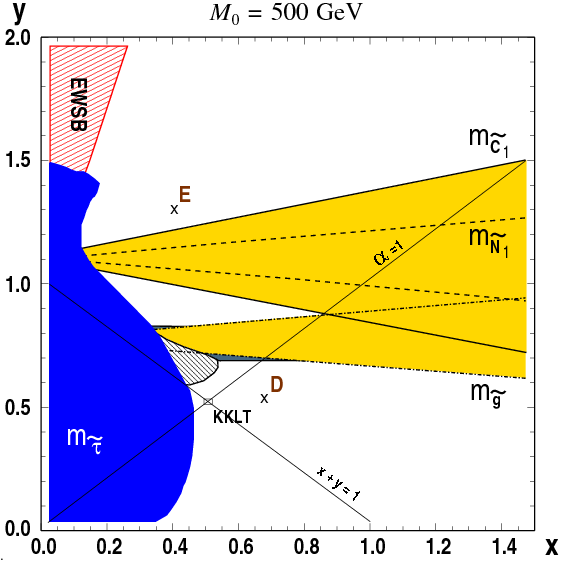}
\includegraphics[scale=0.535]{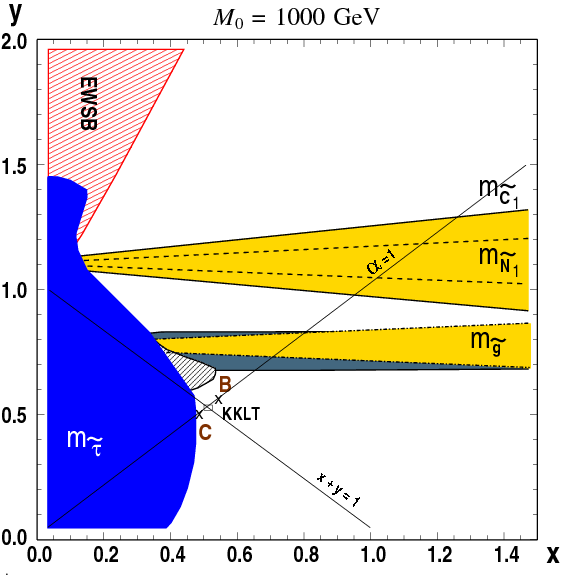}
\includegraphics[scale=0.535]{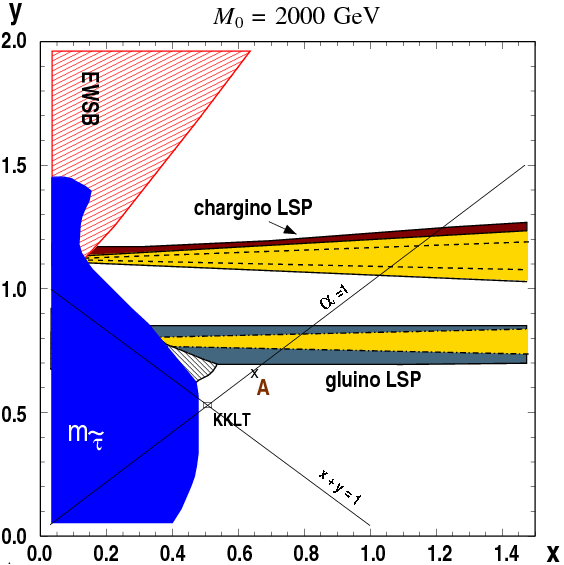}
\caption{\label{fig:param2b}\footnotesize{\textbf{Allowed Parameter
Space for the ``Model-Dependent Scenario.''}} Contours indicate the
locus of points for which $m_{\wtd{N}_1} = 46 \GeV$ (dashed
contour), $m_{\wtd{C}_1} = 103 \GeV$ (solid contour) and
$m_{\wtd{g}} = 200 \GeV$ (dash-dotted contour). The dark shaded
region is the area in which the stau is the LSP, while the smaller
hatched region in the center has a stop LSP for the modular weight
choice of~(\ref{weights}). The hatched region in the upper left of
each plot indicates where no EWSB occurs. For larger values of the
parameter $M_0$ we have indicated the area in which the gluino (or
the chargino) is the LSP by the darker shaded region(s). The labeled
points are the benchmark models of Table~\ref{tbl:models}. The
intersecting lines indicate those points for which $\alpha_m = 1$
(lower left to upper right) and where $R = 0$ (upper left to lower
right). The intersection of these two curves designates the
prediction of the simplest KKLT scenario.}
\end{center}
\end{figure}
%\end{comment}
%=================================================================

In Figure~\ref{fig:param2b} we display the physically allowed region
in the $\lbr x,\,y\rbr$ plane for various values of the mass scale
parameter $M_0$ in the model-dependent case with modular weights
given by~(\ref{weights}). The primary constraint in the $\lbr
x,\,y\rbr$ plane comes from the direct search limits for gauginos.
The mass bounds arising from these searches are somewhat
model-dependent, particularly for the gluino. We will require that
the lightest neutralino have a mass $m_{\wtd{N}_1} \geq 46 \GeV$ and
that the lightest chargino mass obeys $m_{\wtd{C}_1} \geq 103
\GeV$~\cite{Amsler:2008zzb}. For the gluino we will be much more
conservative. Though recent searches at the Tevatron have become
increasing model-independent, we do not wish to rule out too much of
the parameter space of this model class prematurely. We will
therefore require only that $m_{\wtd{g}} \geq 200
\GeV$~\cite{Affolder:2001tc} in our figures, though a bound of
$m_{\wtd{g}} \geq 300 \GeV$~\cite{:2007ww} will be more applicable
over many regions of the parameters. The lightly shaded region is
ruled out by these direct experimental constraints. In addition
there is some parameter space for which the gluino (and for very
large $M_0$, the chargino) can be the LSP, indicated by the darker
shading. For this choice of modular weights the stau is the LSP for
smaller values of $x$ and $y$, indicated by the dark shaded area.
There is also a small region where the lightest stop can be the LSP,
indicated by the hatched region near the center of each figure.
Finally in the upper left corner of each panel there is a region for
which $m_A^2<0$ for the pseudoscalar Higgs and no electroweak
symmetry breaking occurs.

%---------------- Benchmark Table --------------------
\begin{table}[t]
\begin{center}
\begin{tabular}{|c||c|c|c|c|c|c|} \hline
 & Model~A & Model~B & Model~C & Model~D & Model~E & Model~F \\
\hline
$M_0$ & 2000 GeV & 1000 GeV & 1000 GeV & 500 GeV & 500 GeV & 933 GeV \\
$\mgrav$ & 62.6 TeV & 32.0 TeV & 32.0 TeV & 13.2 TeV & 51.3 TeV & 31.7 TeV \\
$\mu_{\rm mess}$ & $10^{12}$ GeV & $10^{8}$ GeV & $10^{12}$ GeV & $10^{10}$ GeV
& $10^{10}$ GeV & N/A \\
\hline
$\alpha_m$ & 1 & 1 & 1 & 0.81 & 3.26 & 1 \\
$\alpha_g$ & 1 & -1/2 & -1 & 0.14 & 1.47 & 0 \\
$N_m$ & 3 & 3 & 3 & 3 & 3 & 0 \\ \hline
$x$ & 0.668 & 0.561 & 0.472 & 0.674 & 0.401 & 0.5 \\
$y$ & 0.668 & 0.561 & 0.472 & 0.543 & 1.307 & 0.5 \\ \hline
\end{tabular}
\end{center}
{\caption{\label{tbl:models}\footnotesize {\bf Benchmark Models}.
The relevant mass scales are given in the first three entries, with
the parameters of Everett et al. in the second block. The final
block re-casts these parameters in terms of the parameterizations of
Choi et al. from~(\ref{xydef}). All models are defined with positive
value of $\mu$ and $\tan\beta=10$. Models~A-B were considered
in~\cite{Everett:2008qy} while Model~C was considered
in~\cite{Everett:2008ey}. Model~F is a mirage model without
messengers near the prediction for the basic KKLT model.}}
\end{table}
%------------------------- END OF THE TABLE ---------------------

This choice of parameter space offers a distinct advantage: the
various theoretical model limits can easily be defined in terms of
the dimensionless quantities $x$ and $y$. In the $\lbr x,\,y\rbr$
plane cases for which $R = 0$ lie along the line for which $x+y=1$.
The line extending from the origin defined by $x=y$ is the case with
$\alpha_m = 1$ for arbitrary $\alpha_g$. The intersection of these
two lines is the simplest prediction of the KKLT framework which
inspired the mirage meditation scenario.
We have imposed these theoretical model lines on top of the allowed
space in Figure~\ref{fig:param2b}. The gauge-mediated limit in
Figure~\ref{fig:param2b} is formally at the origin of the $\lbr
x,\,y\rbr$ plane where $R^{-1} \to 0$, but this limit must be taken
while keeping the product $R\, M_0$ finite~\cite{Choi:2009jn} and is
therefore impossible to reach in our figure. So too, the pure
anomaly-mediated limit is impossible to reach since this is formally
the point $\lbr 0,\,1\rbr$ ({\em i.e.} $\alpha_m \to \infty$) where
the product $\alpha_m\, M_0$ is held fixed. The pure modulus (or
gravity) mediated limit is the point $\lbr 1,\,0\rbr$ in the
figures. For large enough values of $M_0$ there is an expansive
parameter space beyond the mirage mediation ``frontier'' for which a
compressed spectrum of gauginos occurs. The marked points in
Figure~\ref{fig:param2b} represent specific parameter choices which
we will single out for particular study later in the work. The
parameter sets which define these cases are given in
Table~\ref{tbl:models}.

%=(2)============ LSP Properties: Model-Dependent ===============
%\begin{comment}
\begin{figure}[t]
\begin{center}
\includegraphics[scale=0.535]{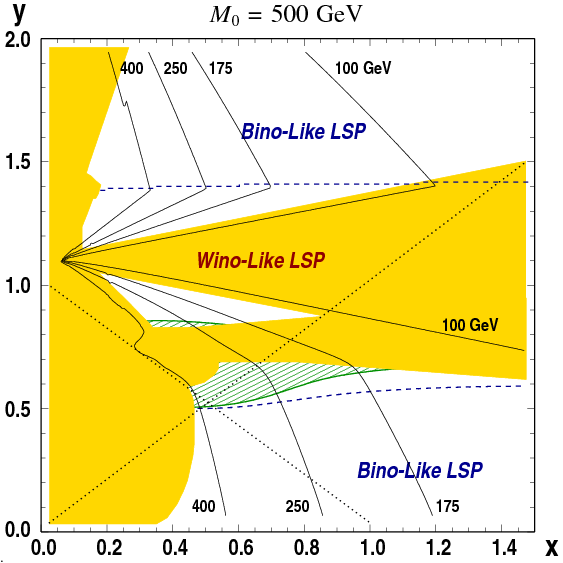}
\includegraphics[scale=0.535]{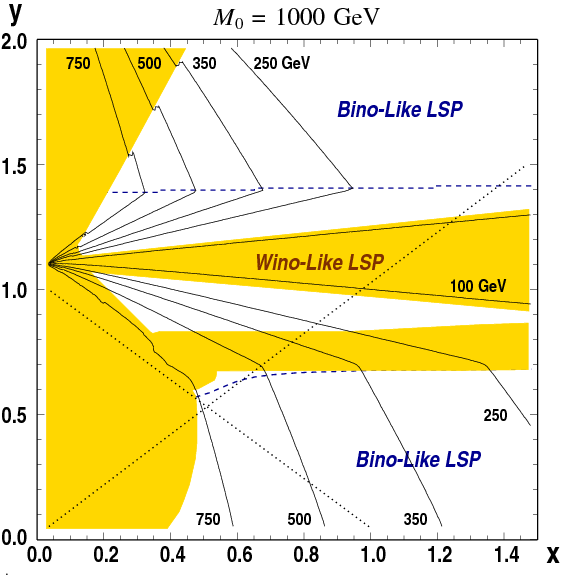}
\caption{\label{fig:LSP2b}\footnotesize{\textbf{LSP Properties for
the ``Model-Dependent Scenario.''}} Contours of constant LSP mass
$m_{\tilde{N}_1}$ are given for the scenario with modular weights
of~(\ref{weights}) for mass scale $M_0 =500\GeV$ (left panel) and
$M_0 =1000\GeV$ (right panel). The lightly shaded region is ruled
out for the reasons indicated in Figure~\ref{fig:param2b}. For both
values of $M_0$ the LSP is primarily bino-like for $y\lappeq 0.6$
and $y \gappeq 1.4$. For the case of $M_0 = 500\GeV$ there is also
some parameter space where the LSP is a mixture of Higgsino and
wino, indicated by the hatched region.}
\end{center}
\end{figure}
%\end{comment}
%=================================================================

In Figure~\ref{fig:LSP2b} we concentrate on the properties of the
lightest neutralino for the specific cases of $M_0 = 500 \GeV$ (left
panel) and $M_0 = 1000 \GeV$ (right panel). The union of all the
theoretically and experimentally forbidden regions in
Figure~\ref{fig:param2b} is indicated by the light shading. Note
that the abrupt kinks in the contours of constant $m_{\tilde{N}_1}$
are the result of level crossings where the neutralino goes from
being over 95\% wino-like (along $y\simeq 1$) to over 95\% bino-like
($y\gappeq 1.4$ and $y\lappeq 0.6$). The region with a wino-like LSP
is indicated in Figure~\ref{fig:LSP2b} by the nearly horizontal
dashed lines. For the case of $M_0 = 500\GeV$ there is also some
parameter space where the LSP is a mixture of Higgsino and wino,
indicated by the hatched region. The analogous region for the case
of $M_0 = 1000\GeV$ would lie within the phenomenologically
forbidden region.

%=(3)============ Sparticle Masses: Model-Dependent ===============
%\begin{comment}
\begin{figure}[p]
\begin{center}
\includegraphics[scale=0.35]{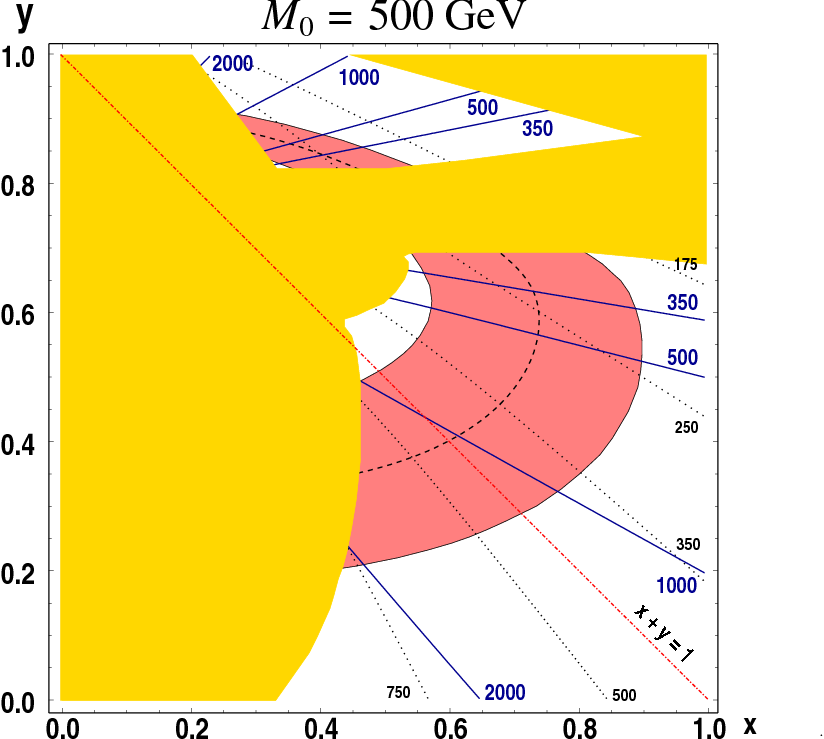}
\includegraphics[scale=0.35]{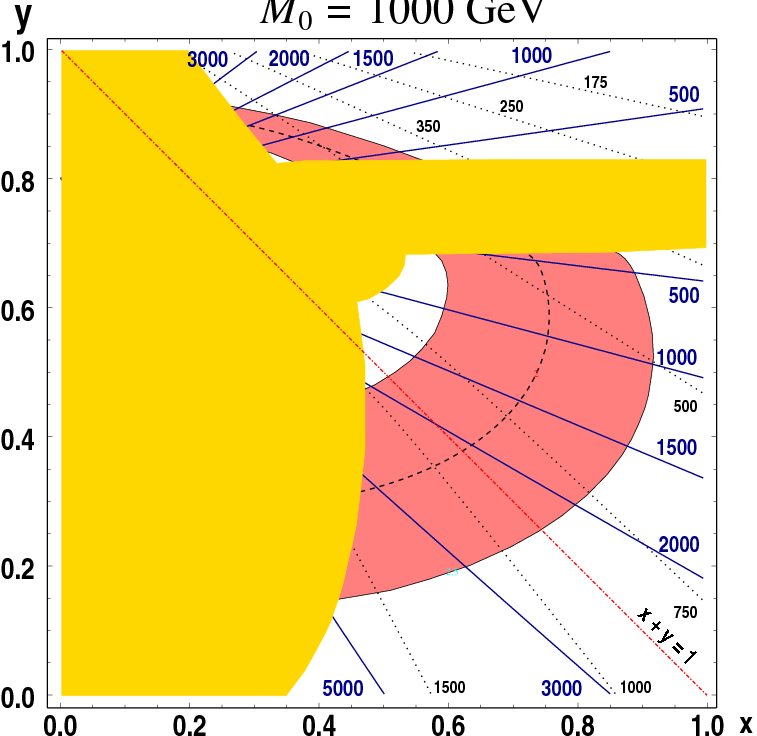}
\includegraphics[scale=0.35]{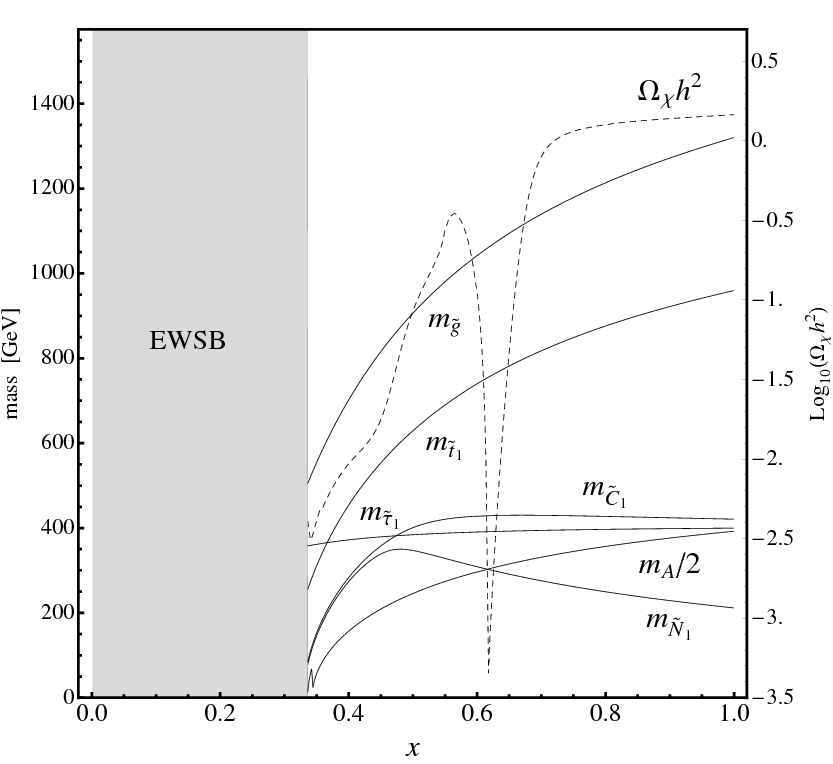}
\includegraphics[scale=0.35]{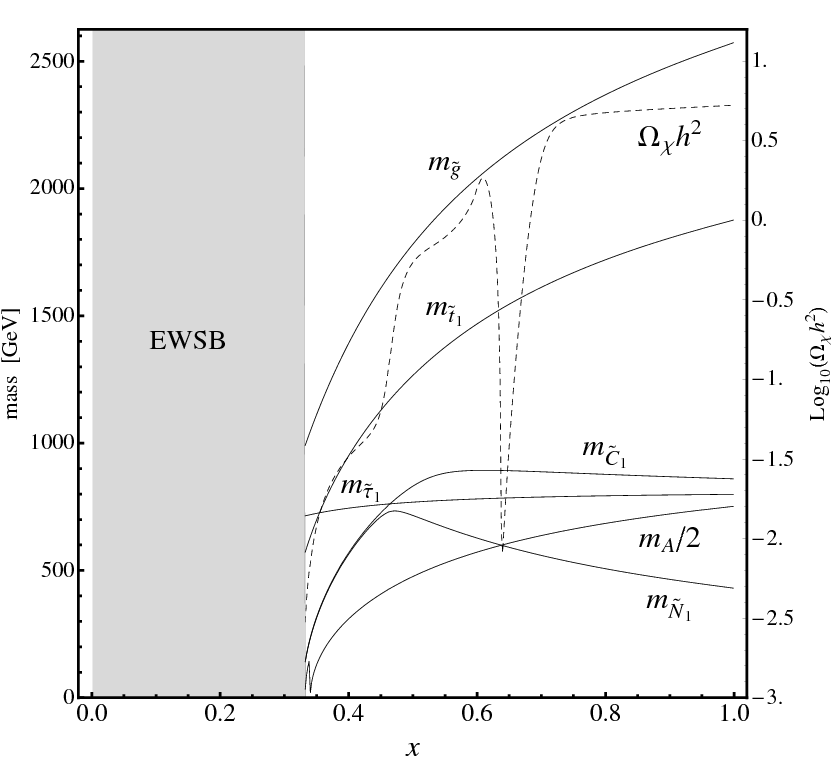}
\caption{\label{fig:masses2b}\footnotesize{\textbf{Key Superpartner
Masses for the ``Model-Dependent Scenario.''}} Upper panels give
various masses for the $\lbr x,\, y \rbr$ plane, while the lower
panels plot masses along the line $x+y =1$ with $N_m =0$. Panels on
the left take $M_0 = 500\GeV$ while those on the right take $M_0 =
1000\GeV$. The lightly shaded region in all plots is
phenomenologically forbidden. The darker shaded region in the upper
plots shows the area where $1.5 \leq m_A/m_{\tilde{N}_1} \leq 2.5$,
with $m_A = 2m_{\tilde{N}_1}$ given by the curved dashed line. In
the upper plot heavy solid lines are contours of constant
$m_{\tilde{g}}$ while dashed lines are contours of constant
$m_{\tilde{C}_1}$. All masses are in GeV.}
\end{center}
\end{figure}
%\end{comment}
%=================================================================

Some of the remaining superpartner masses are given in
Figure~\ref{fig:masses2b}, again for the cases of $M_0 = 500 \GeV$
(left panel) and $M_0 = 1000 \GeV$ (right panel). As before, the
lightly shaded region is excluded. The upper panels show contours of
constant gluino mass (solid heavy lines) and constant (lightest)
chargino mass (dashed lines) for the entire $\lbr x,\, y \rbr$
plane. The curved shaded region represents the area where $1.5 \leq
m_A/m_{\tilde{N}_1} \leq 2.5$. In this region we expected enhanced
annihilation of relic neutralinos though the pseudoscalar Higgs
resonance -- particularly along the curved dashed contour for which
$m_A = 2 m_{\tilde{N}_1}$. Note that we have restricted our
attention to the region in which both $x$ and $y$ lie between zero
and one. This is the range explored in the mirage family of models.
In the lower two panels we plot several superpartner masses as a
function of the parameter $x$ where we take $N_m = 0$, {\em i.e.} a
true ``mirage'' model. The phenomenology in the lower panels roughly
corresponds to following the line $x+y=1$ in the upper two panels.
Note that the correspondence between the plots, while very good, is
not precise. Strictly speaking, the line $x+y=1$ is determined by
the requirement that $R=1$. But from the definition in~(\ref{Rdef})
it is apparent that there are two solutions for this constraint --
one is the mirage limit where $N_m = 0$. If we fix the number of
messengers at a value $N_m \neq 0$ (as in the case of the top panels
in Figure~\ref{fig:masses2b}) then the line $x+y=1$ forces the
second solution in which the quantity in braces in~(\ref{Rdef})
vanishes. This is necessarily a solution for which $\alpha_g$ does
not vanish. Nevertheless, the phenomenology is similar to that of
the mirage model near this line, so the reader can use the line
$x+y=1$ as a reasonable proxy for the mirage model limit.

%=(4)============ Allowed Region: Model-Independent ===============
%\begin{comment}
\begin{figure}[t]
\begin{center}
\includegraphics[scale=0.535]{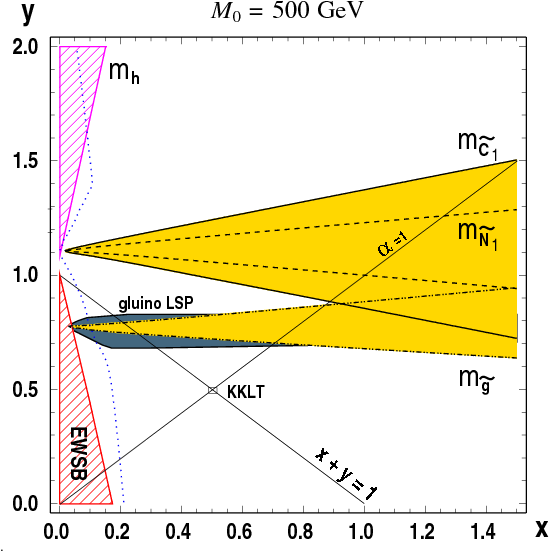}
\includegraphics[scale=0.535]{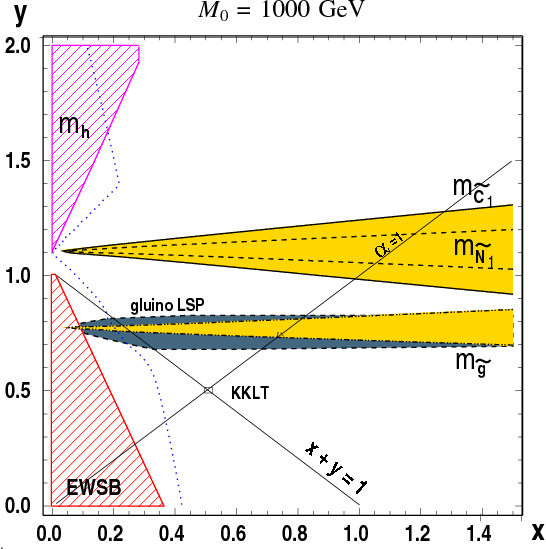}
\caption{\label{fig:param2a}\footnotesize{\textbf{Allowed Parameter
Space for the ``Model-Independent Scenario.''}} Gaugino mass
contours are the same as Figure~\ref{fig:param2b}. The darker shaded
region is the area in which the gluino is the LSP. The dotted line
on the left of each plot is the contour where $m_{\wtd{N}_1} = 1
\TeV$. The hatched region in the lower left of each plot indicates
where $m_A^2 < 0$ and no EWSB occurs for a value $\mu = 1\TeV$. We
have indicated the area in the upper left where $m_h \leq 100 \GeV$
due to large radiative corrections.}
\end{center}
\end{figure}
%\end{comment}
%=================================================================

Finally, we wish to point out that much of the theoretically
forbidden region is an artifact of the choice~(\ref{weights}) for
modular weights. It is clear from~(\ref{AtermUV})
and~(\ref{scalarUV}) that the choice of modular weights will have
$\order(1)$ effects on certain key parameters. This, in turn,
affects the low-scale physical masses and the derived value of the
$\mu$-parameter once electroweak symmetry breaking is imposed. The
amount of the $\lbr x ,\,y \rbr$ plane available is therefore
heavily dependent on these model choices, as was pointed out
in~\cite{Baer:2006tb,Choi:2006im,Baer:2007eh}. We therefore show the
allowed parameter space in Figure~\ref{fig:param2a} for the
``model-independent'' scenario in which we set $\mu = m_A = 1 \TeV$
and set all scalar masses to the value of the gluino soft mass $M_3$
or to 1~TeV, whichever is larger. As $x\to0$ all gauginos get
extremely massive for fixed $M_0$ value. Thus, near the $x=0$ axis
there is always a region where $m_{\wtd{N}_1} = 1\TeV$, indicated by
the dotted contour, where a scalar particle would be the LSP for our
choice of scalar masses. As the gluino gets increasingly massive in
the $x \to 0$ limit it induces large radiative corrections to the
Higgs potential, producing regions where $m_h \leq 100 \GeV$ (upper
left corners) or where $m_A^2 \leq 0$ (lower left corners). The
exact location of these contours depends on the choices made for
$\mu$ and the soft scalar masses.

As we have seen the allowed parameter space depends on the scalar
sector of the theory, which is set by the modular weights assigned
to the multiplets of the MSSM. The discrete choices for modular
weights can be seen as selecting (a) the relative size of the
gaugino masses versus scalar masses and (b) determining the degree
of non-universality in the scalar masses between the Higgs sector
and the matter sector. The former is akin to the choice of $m_0$
versus $m_{1/2}$ in mSUGRA models, while the latter is related to
certain non-universal extensions of these mSUGRA models. In order to
capture this physics, in the next section we will consider the dark
matter signals of the LSP neutralino in terms of the specific
benchmark case of~(\ref{weights}) over the space defined by $\lbr
x,\, y ,\, M_0 \rbr$. But we will also occasionally relax this heavy
constraint and randomize over all other parameters that determine
the model. Together these two strategies will give a reasonable
depiction of the dark matter implications of the gaugino masses
in~(\ref{Ma2}).

\noindent\section{Survey of Dark Matter Signatures}
\label{sec:signals}
\subsection{Thermal Relic Density}

The phenomenology of relic neutralino dark matter consists of two
largely disconnected aspects: the cosmological density of relic
neutralinos and the physics of these neutralinos here in our galaxy
today. These aspects are logically distinct, but can be related in
any comprehensive theory of the origin and nature of dark matter.
The physics of neutralino annihilation, or neutralino interaction
with terrestrial dark matter detectors, depends on not much more
than the properties of the LSP itself: its wave-function and mass.
It also depends on its density in our local halo and certain other
astrophysical properties independent of the nature of the LSP. The
cosmological relic abundance, on the other hand, can be determined
by this information only in certain special cases. More often it
depends on the physical masses of other particles (supersymmetric
states and Higgs states) that were part of the relativistic plasma
at some earlier, hotter time in the cosmos. Crucially, the
calculation of the cosmological relic density depends on certain
assumptions about the thermal history of the universe.

\begin{figure}[t]
\begin{center}
\includegraphics[scale=0.57]{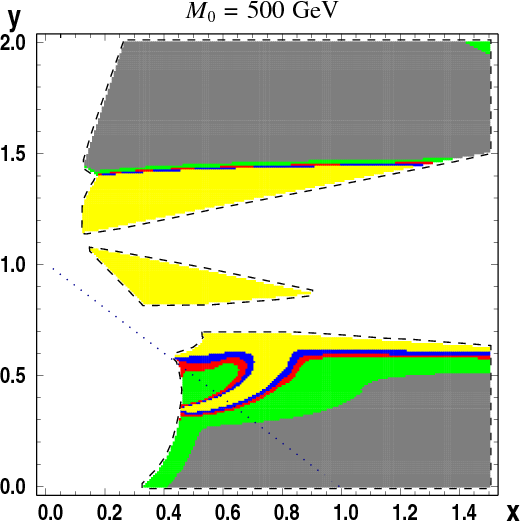}
\includegraphics[scale=0.55]{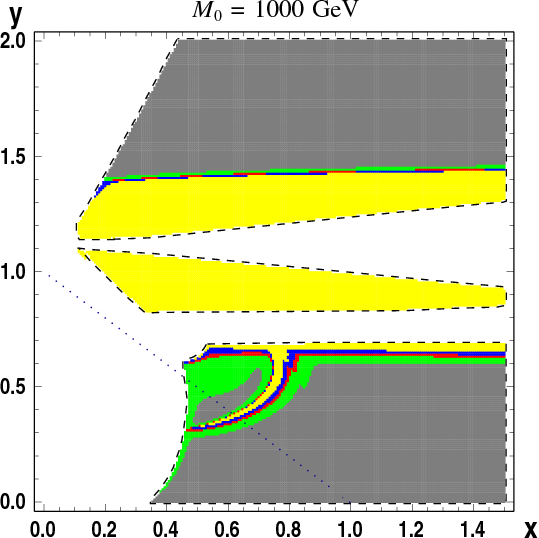}
\caption{\label{fig:p2_relic}\footnotesize{\textbf{Thermal Relic
Density in Deflected Mirage Mediation}.} Left panel takes $M_0 =
500\GeV$, right panel takes $M_0 = 1000\GeV$. The ``WMAP preferred''
region of $0.07 \leq \Ochi \leq 0.14$ is here indicated by the
narrow red shaded region. For the wino-like LSP and within the
$A$-funnel region the relic density drops below the critical value
of $\Ochi = 0.025$, indicated by the yellow (very light) shading.
The remaining regions are $0.025 \leq \Ochi \leq 0.07$ (blue), $0.14
\leq \Ochi \leq 1$ (green) and $\Ochi >1$ (gray).}
\end{center}
\end{figure}

The most important inputs to this thermal relic abundance
calculation are the mass of the LSP and the mass difference between
it and the next lightest superpartner. The thermal relic abundance
of the LSP neutralino for the theory with modular
weights~(\ref{weights}) is given in Figure~\ref{fig:p2_relic} as a
function of the dimensionless parameters~$x$ and~$y$ for the cases
$M_0 = 500 \GeV$ (left panel) and $M_0 = 1000 \GeV$ (right panel).
Here and throughout, all calculations are performed using the
computer package {\tt DarkSUSY 5.0.4}~\cite{Gondolo:2004sc} after
computing the physical mass spectrum throughout the parameter space
using {\tt SuSpect 2.4}. The WMAP three-year
data~\cite{Spergel:2006hy} is best fit by a relic density in the
range
\beq 0.0855 \leq \Ochi \leq 0.1189\, , \label{omegah2} \eeq
at the $2\sigma$ level. In our figures we will expand this region
somewhat to a ``WMAP preferred'' region
\beq 0.07 \leq \Ochi \leq 0.14\, , \label{prefer} \eeq
which will be slightly easier to resolve than the very narrow band
in~(\ref{omegah2}).

%=(5)============ Parameter Set 2: Relic Density ===============
%\begin{comment}
%\end{comment}
%=================================================================

In Figure~\ref{fig:p2_relic} this ``WMAP preferred'' region is
indicated by the narrow red shaded region. Inside the parameter
space with a wino-like LSP, and within the $A$-funnel region, the
relic density drops below the value of $\Ochi = 0.025$, indicated by
the yellow shading. The intermediate regime of $0.025 \leq \Ochi
\leq 0.07$ is indicated by blue shading. The green area in the plot
has $0.14 \leq \Ochi \leq 1$ while the gray area has $\Ochi > 1$.
The transition from heavily wino-like LSP to heavily bino-like LSP
occurs near $y\simeq 1.4$ where the relic density rapidly changes
from far too low to far too high. In between these extremes there
exists a narrow region with $\Ochi \simeq 0.10$.

The precision with which the cosmological abundance of cold dark
matter can be inferred from the cosmic microwave background is
remarkable. Yet despite the narrow window of~(\ref{omegah2}) we
cannot say that all of this cold dark matter is composed of
neutralinos, nor that it was produced by standard thermal
mechanisms. Thus we must be careful not to immediately exclude those
parameter sets for which the thermal calculation yields $\Ochi$
outside the range of~(\ref{omegah2}). For example, an
under-abundance of relic neutralinos may simply indicate the
presence of non-thermal production mechanisms for stable
LSPs~\cite{Moroi:1994rs,Kawasaki:1995cy,Moroi:1999zb}. Such
mechanisms are especially well-motivated in string-inspired contexts
such as in the KKLT model
framework~\cite{Nagai:2007ud,Nagai:2008se}. An over-abundance of
relic neutralinos is a greater cause for concern, but here too
mechanisms exist which can bring the current relic density in line
with the results from
WMAP~\cite{Gelmini:2006pw,Gelmini:2006pq,Lahanas:2006xv}. We prefer
to remain agnostic on the issue as this is not the physics of
interest to us. In fact, the physics of relic neutralino detection
can be separated from the cosmological density of neutralinos to the
extent that it is only the {\em local} halo density profile that is
relevant. Of course the two will be related once a set of
assumptions about the physics of the early universe are specified.
We will, however, make one exception: for cases with $\Ochi \leq
0.025$ (indicated by the yellow shading in
Figure~\ref{fig:p2_relic}) it becomes difficult for the relic
particle in question to account adequately for the local halo
density of our galaxy~\cite{Jungman:1995df}. Therefore when
calculating observable quantities that depend on the relic
neutralino number density $n_{\chi}$ present in our galaxy (or the
energy density $\rho_{\chi} = m_{\chi} n_{\chi}$) we will rescale
the assumed local density of $(\rho_{\chi})_0 = 0.3$~GeV/cm$^3$ by
the multiplicative factor $r_{\chi} = {\rm Min}(1,\, \Ochi/0.025)$.

%---------------- Scatter Table --------------------
\begin{table}[t]
\begin{center}
\begin{tabular}{|c||c|c||c||c|c|} \hline
Parameter & Min Value & Max Value & Parameter & Min Value & Max Value\\
\hline
$M_0$ & 100 GeV & 2000 GeV & $\mu_{\rm mess}$ & $10^8$ GeV & $10^{14}$ GeV \\
$\alpha_m$ & 0 & 5 & $\alpha_g$ & -5 & 5 \\
$\tan\beta$ & 2 & 40 & $N_m$ & 1 & 3 \\
%
%$n_m$ & 0 & 1/2 & $n_H$ & 0 & 1 \\
\hline
\end{tabular}
\end{center}
{\caption{\label{tbl:scatter}\footnotesize {\bf Parameter Scan
Ranges for Deflected Mirage Mediation Models}. In an effort to
average over possible model-dependent effects we have produced 1000
deflected mirage mediation models by randomly choosing input
parameters from within the ranges indicated. An additional random
choice of modular weights for the matter and Higgs multiplets was
also made to complete the model.}}
\end{table}
%------------------------- END OF THE TABLE ---------------------

Displaying results in the $\lbr x,\, y\rbr$ plane defined
by~(\ref{xydef}) is useful for making contact with certain
theoretical limits, but requires fixing a number of important input
parameters. We will therefore also show results of a scan over
parameter inputs for the deflected mirage mediation paradigm. The
quantities we vary are listed in Table~\ref{tbl:scatter}. Note that
the input parameters $\Lambda_{\rm mess}$ and $m_{3/2}$ are obtained
from the definitions in~(\ref{alphag}) and~(\ref{alpham}),
respectively. For the modular weights we randomly chose from the
possibilities
\begin{eqnarray} n_Q = n_U = n_D = n_L = n_E &=& 0\,\,{\rm
or}\,\,1/2  \nonumber \\
n_{H_u} = n_{H_d} &=& 0\,\,{\rm or}\,\,1/2\,\,{\rm or}\,\,1
\label{nrandom} \end{eqnarray}
which include the cases with the widest range of allowed points at
the electroweak scale~\cite{Baer:2007eh}. We generated 1000~points,
all of which have proper electroweak symmetry breaking and satisfy
the mass bounds imposed in Section~\ref{sec:space}. For the sake of
comparison we also generated 1000~points with $\alpha_m = \alpha_g =
0$ and vanishing modular weights. These points represent unified
modulus-mediated models with a phenomenology similar to traditional
dilaton-dominated models~\cite{Barbieri:1993jk,Brignole:1993dj}.

%=(6)============ Scatter: Relic density ========
%\begin{comment}
\begin{figure}[t]
\begin{center}
\includegraphics[scale=0.75]{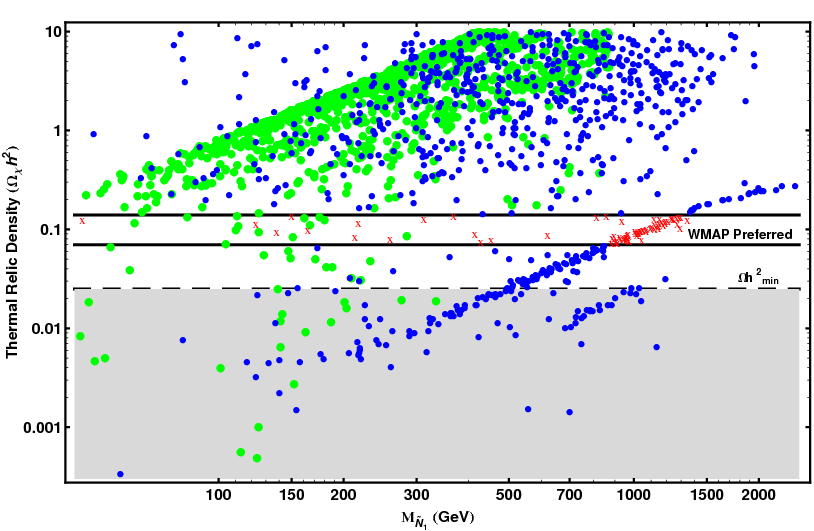}
\caption{\label{fig:relic}\footnotesize{\textbf{Relic Density for
Models from Table~\ref{tbl:scatter}.}} Thermal relic density $\Ochi$
is plotted as a function of the neutralino LSP mass. Dark circles
are the ensemble of deflected mirage mediation models defined by
Table~\ref{tbl:scatter}. Larger lighter circles are an ensemble of
unified modulus-mediated models with $\alpha_m = \alpha_g = 0$. Red
crosses are the DMM models which fall into the ``WMAP Preferred''
region $0.07 < \Ochi < 0.14$. Models in the shaded region below
$\Omega\,{\rm h}_{\rm min}^2 = 0.025$ have the local halo density
rescaled before any further observable is calculated.}
\end{center}
\end{figure}
%\end{comment}
%=================================================================

We plot the thermal relic density $\Ochi$ as a function of the LSP
mass for both the mSUGRA-like case $\alpha_m = \alpha_g = 0$ and for
the ensemble of deflected mirage mediation models defined by
Table~\ref{tbl:scatter} in Figure~\ref{fig:relic}. Dark circles are
the ensemble of deflected mirage mediation models, while the larger,
lighter circles are the cases with $\alpha_m = \alpha_g = 0$. Red
crosses are the DMM models which fall into the ``WMAP preferred''
region~(\ref{prefer}). Note that the WMAP-preferred area for the DMM
model set tends to cluster around $m_{\tilde{N}_1} \sim
\order(1\TeV)$, as opposed to the unified models which require a
much lower-mass LSP. We will designate this group of models with red
crosses in the figures which follow. Finally, we note that those
model in the shaded region below $\Omega\,{\rm h}_{\rm min}^2 =
0.025$ have the local halo density rescaled by the factor $r_{\chi}
= {\rm Min}(1,\, \Ochi/0.025)$ before any further observable is
calculated.

%%%%%%%%%%%%%%%%%%%%%%%%%%%%%%%%%%%%%%%%%%%%%%%%%%%%%%%%%%%%%
%%%%%%%%%%%%%%% DIRECT DETECTION %%%%%%%%%%%%%%%%%%%%%%%%%%%%
%%%%%%%%%%%%%%%%%%%%%%%%%%%%%%%%%%%%%%%%%%%%%%%%%%%%%%%%%%%%%
\subsection{Direct Detection}

Given the various theoretical considerations outlined above, it is
clear that terrestrial direct detection experiments would provide
the firmest evidence for the existence of weakly-interacting massive
dark matter. The interaction rate of relic neutralinos with target
nuclei depends on the local density of the neutralino in the local
halo. This number is more tightly constrained by experimental
observation~\cite{Jungman:1995df,Bergstrom:1997fj} and fits to large
scale structure simulations of galaxies similar to ours. The rate
depends heavily on the properties of the neutralino itself, but it
also requires the input of certain nuclear matrix elements that must
be inferred from experiment. The values of these matrix elements are
uncertain -- often to a surprisingly large degree. The resulting
uncertainty in the interaction cross-section can be as much as 50\%
in some circumstances~\cite{Barger:2007nv,Ellis:2008hf}.
Though our interest here is in the study of broad correlations
between the parameters of the generalized mediation scenario with
the properties of the relic neutralino, we should keep these
uncertainties in mind when discussing the results of any
calculation.

% ------------- Direct Detection Experiments ------------------
%
\begin{table}[t]
\begin{center}
\begin{tabular}{|c|l||c|c||c|} \hline
Ref. & Experiment Name & Fiducial Mass [kg] & Exposure Time [yr]
& $R_{10}$ [counts/(kg yr)] \\
\hline
%\begin{center}
%\begin{table}[h!]
%\begin{tabular}{c|c|c|c|c}
%experiment & target & fiducial mass [kg] & time [yr] & rate for 10
%events [counts/(kg yr)]\\ \hline
\cite{Angle:2007uj} & XENON10 & 5.4 & 0.16 & 11.54 \\
\cite{Aprile:2004ey} & XENON100 & 170 $\times$ 0.8 & 1 & $7.35\times 10^{-2}$ \\
\cite{LUX} & LUX & 350 $\times$ 0.8 & 3 & $1.19\times 10^{-2}$ \\
\cite{Aprile:2004ey} & XENON1T & 1000 $\times $0.8 & 5 & $2.50\times 10^{-3}$ \\
\hline
\cite{Ahmed:2008eu} & CDMS II & 3.75 & 0.29 & 9.18 \\
\cite{Akerib:2006rr} & SuperCDMS (SNOlab) & 27 $\times$ 0.8 & 3 & 0.15 \\
\cite{Akerib:2006rr} & SuperCDMS (DUSEL) & 1140 $\times$ 0.8 & 5 &
$2.19\times 10^{-3}$ \\
\hline
\end{tabular}
\end{center}
{\caption{\label{table:ddrates}\footnotesize {\bf Rate Estimates for
Various Experiments}. The minimum threshold rate $R_{10}$ necessary
to produce 10 events in a given experiment for the fiducial mass and
exposure time given is tabulated in the final column. Note that for
Xenon10 and CDMS~II we use the experimentally quoted fiducial
masses. For all other (future) experiments we assume a fiducial mass
equivalent to 80\% of the nominal quoted target mass.}}
\end{table}
%---------------------------------------------------------

We will focus here exclusively on two types of detector design:
cryogenic germanium bolometers and dual-phase liquid/gas xenon
detectors. These types are currently operational in at least one
experiment and producing results -- and future enlargements are
envisioned for each. We use {\tt DarkSUSY} with default settings for
all nuclear form factors to obtain the differential rate of
interactions per unit recoil energy on germanium and
xenon.\footnote{We have corrected a normalization error in the
recoil rate routines and ensured that all results are rescaled (when
necessary) by the thermal relic density of LSPs.} We do so over a
range of recoil energies relevant to the desired experiment. As
in~\cite{Altunkaynak:2008ry}, we perform the integration of these
rates using two possible energy ranges
\bea {\rm Xe} &:& 5\,{\rm keV} \leq E_{\rm recoil} \leq 25\,{\rm
keV} \nonumber \\ {\rm Ge} &:& 10\,{\rm keV} \leq E_{\rm recoil}
\leq 100\,{\rm keV} \, . \label{rates} \eea
Note that these rate calculations include both spin-independent and
spin-dependent contributions.

%=(7)============ Parameter Set 3: Direct Detection (Xe) ===============
\begin{figure}[p]
\begin{center}
\includegraphics[scale=0.52]{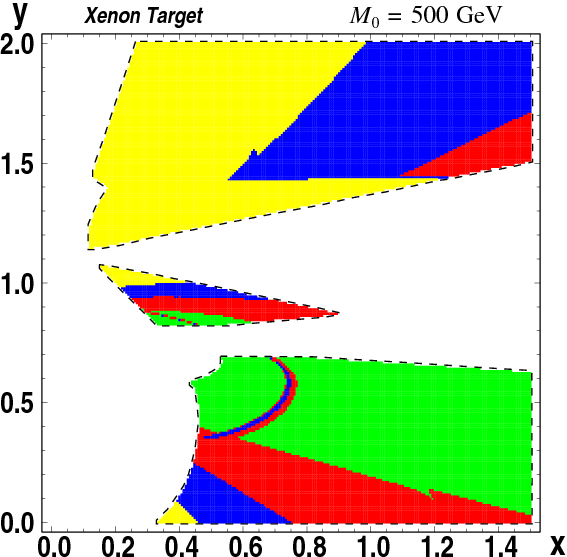}
\includegraphics[scale=0.52]{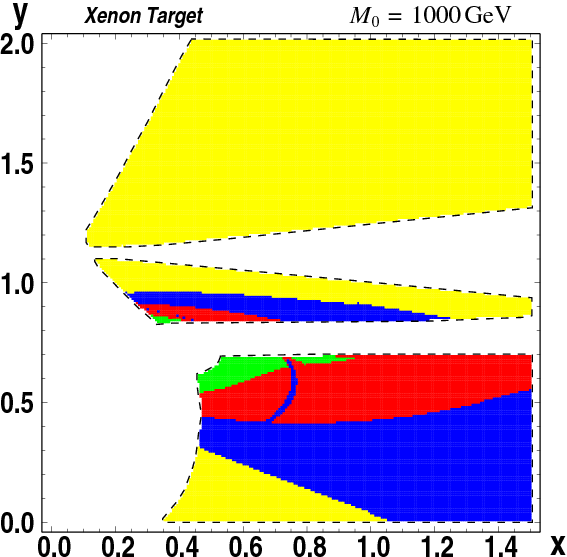}
\includegraphics[scale=0.52]{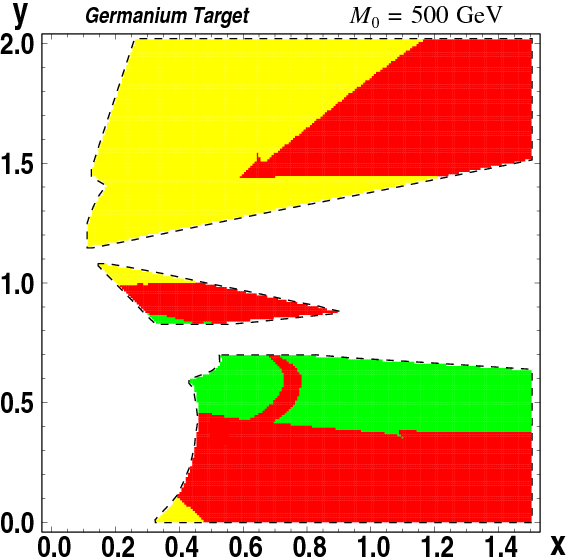}
\includegraphics[scale=0.52]{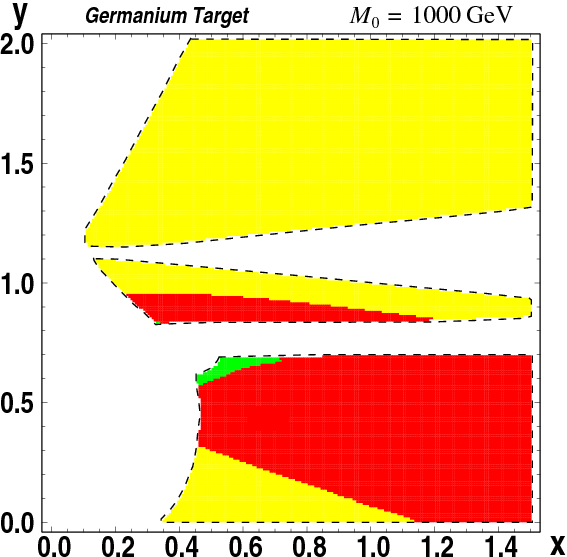}
\caption{\label{fig:dd}\footnotesize{\textbf{Neutralino Recoil Rates
on Xenon (top) and Germanium (bottom).}} Left panels set $M_0 =
500\GeV$, right panels set $M_0 = 1000 \GeV$. Phenomenologically
allowed areas are enclosed by the heavy dashed lines. Colored
shading indicates the reach of future direct detection experiments,
as computed in Table~\ref{table:ddrates}. Yellow in all panels
indicates parameter space that will be inaccessible to any of the
future experiments in Table~\ref{table:ddrates}. For xenon targets,
green indicates $11.54 > R^{\rm Xe}_{10} \geq 0.0735$, red indicates
$0.0735 > R_{10}^{\rm Xe} \geq 0.0119$ and blue indicates $0.0119 >
R_{10}^{\rm Xe} \geq 0.0025$ in recoils per kg-year. For germanium
targets, green indicates $9.18 > R^{\rm Ge}_{10} \geq 0.15$ and red
indicates $0.15 > R^{\rm Ge}_{10} \geq 0.00219$ recoils per
kg-year.}
\end{center}
\end{figure}
%\end{comment}
%=================================================================

Bounds on these recoil rates have been set by current direct search
experiments such as CDMS~II \cite{Ahmed:2008eu} and
Xenon10~\cite{Angle:2007uj}, using exposures of 316.4 kg-days on
xenon and 397.8 kg-days on germanium, respectively. Expected
backgrounds for both experiments are low, typically on the order of
(or less than) 10~expected background events per year of exposure.
Background rates at future enlargements are expected to be even
lower. We will therefore compute the necessary interaction rates
over the energy ranges in~(\ref{rates}) to produce 10 events for a
given mass and exposure time in a number of current or planned
experiments based on xenon or germanium targets. We summarize the
detector details we will assume in Table~\ref{table:ddrates}. For
the current experiments a limit can be placed by requiring $R\leq
R_{10}$, the rate for which 10 events would have been produced at
each experiment for the quoted exposure. These numbers are given in
the final column of Table~\ref{table:ddrates} and indicated in
Figure~\ref{fig:dd} by the various shaded regions. We find that none
of the model points give rates which should have been detected in
the initial runs of Xenon10 and CDMS~II.

\begin{figure}[t]
\begin{center}
\includegraphics[scale=0.52]{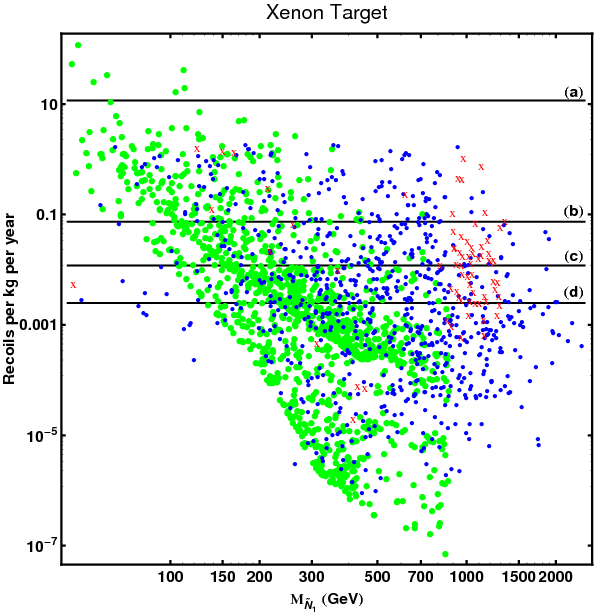}
\includegraphics[scale=0.52]{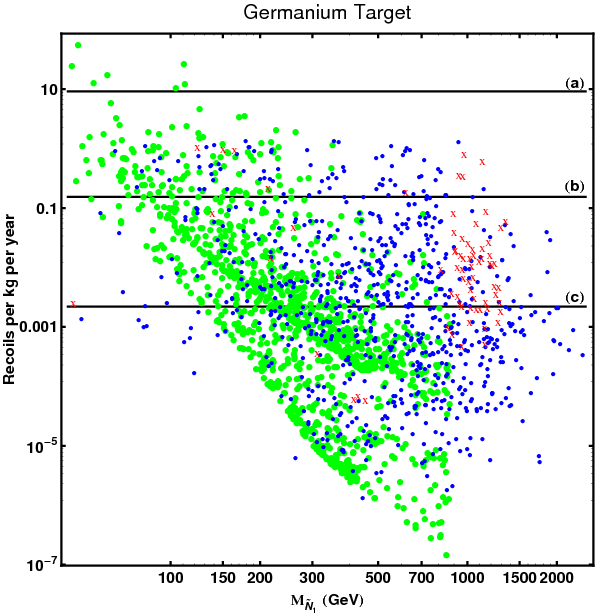}
\caption{\label{fig:directXeGe}\footnotesize{\textbf{Recoil Rates in
Direct Detection Experiments for Models from
Table~\ref{tbl:scatter}}.} Reach of current and future experiments
in terms of the $R_{10}$ values of Table~\ref{table:ddrates} are
given for xenon-based experiments (left panel) and germanium-based
experiments (right panel). Dark circles are the ensemble of
deflected mirage mediation models defined by
Table~\ref{tbl:scatter}. Larger lighter circles are an ensemble of
unified modulus-mediated models with $\alpha_m = \alpha_g = 0$. Red
crosses are the DMM models which fall into the ``WMAP preferred''
region~(\ref{prefer}). The xenon-based experiments are (a) Xenon10
(reported), (b) Xenon100 (1 year), (c) LUX (3 year) and (d) Xenon1T
(5 year). The germanium-based experiments are (a) CDMS~II
(reported), (b) SuperCDMS (SNOLab - 3 year) and (c) SuperCDMS (DUSEL
- 5 year).}
\end{center}
\end{figure}

For the future experiments listed in Table~\ref{table:ddrates} we
computed the value of $R_{10}$ assuming a fiducial target mass equal
to 80\% of the quoted (nominal) mass and the exposure time listed.
The results (in units of recoils per kg per year) for the $\lbr
x,\,y\rbr$ plane are plotted in Figure~\ref{fig:dd} for xenon and
germanium targets, respectively. Results for our random model survey
are given in Figure~\ref{fig:directXeGe}. The yellow (light shading)
in all panels indicates parameter space that will be inaccessible to
any of the future experiments in Table~\ref{table:ddrates}. At
roughly the 100 kg-year level of exposure a fair amount of the
parameter space will be detectable, particularly for low mass
scales. This is indicated by the green (medium shading) in all
panels, for which $11.54 > R^{\rm Xe}_{10} \geq 0.0735$ counts/kg-yr
and $9.18 > R^{\rm Ge}_{10} \geq 0.15$ counts/kg-yr for xenon and
germanium, respectively. This is within reach of the Xenon100
experiment after about one year of operation, or the SuperCDMS
SNOLab proposal after about three years of operation. In about one
ton-year of exposure -- equivalent to about three years of the LUX
experiment -- the red (dark) shaded region in the xenon panels with
$0.0735 > R_{10}^{\rm Xe} \geq 0.0119$ counts/kg-yr will be probed.
Finally, after approximately five ton-years in xenon the blue shaded
region with $0.0119 > R_{10}^{\rm Xe} \geq 0.0025$ counts/kg-yr can
be explored, or the red shaded region with $0.15 > R^{\rm Ge}_{10}
\geq 0.00219$ counts/kg-yr in germanium. Note that these regions are
related to the wave-function of the neutralino LSP and to the
presence of light Higgs states in the spectrum. In the $A$-funnel
region where the relic density is well below the $\Ochi = 0.025$
threshold the interaction rates have been rescaled, thereby
requiring a much larger exposure for detection.

%=(8)============ Scatter: Direct Detection (Xe) ========
%\begin{comment}
%\end{comment}
%=================================================================

The results of the survey of randomized models are given in
Figure~\ref{fig:directXeGe}. The current experimental limits are
given by the horizontal lines labeled~(a) in the figures. Though a
few of our unified models would have been detected at these
experiments, none of the deflected mirage models would have produced
at least ten recoils. The five ton-year limit is indicated by the
horizontal lines labeled~(d) for xenon (with $R_{10}^{\rm Xe} =
2.50\times 10^{-3}$ counts/kg-yr) and~(c) for germanium (with
$R_{10}^{\rm Ge} = 2.19\times 10^{-3}$ counts/kg-yr). The majority
of the points which satisfy the WMAP preferred bound
of~(\ref{prefer}) can be probed at this level of exposure.

%%%%%%%%%%%%%%%%%%%%%%%%%%%%%%%%%%%%%%%%%%%%%%%%%%%%%%%%%%%%%
%%%%%%%%%%%%%%% INDIRECT: MUONS %%%%%%%%%%%%%%%%%%%%%%%%%%%%%
%%%%%%%%%%%%%%%%%%%%%%%%%%%%%%%%%%%%%%%%%%%%%%%%%%%%%%%%%%%%%
\subsection{Muons}

The presence of relic neutralinos can also be inferred from
experiments which seek to detect the products of neutralino
annihilation processes. The rate for neutralino pair annihilation
will be quite low except in those areas where the present-day relic
density is high. For indirect detection searches to be successful,
therefore, they must be sensitive to objects that (a) are able to
travel unimpeded for long distances from their point of origin, and
(b) are not themselves copiously produced by mundane astrophysical
processes. For the second point, it may be sufficient to demand that
some property of the object (such as the differential flux as a
function of energy) be significantly different from estimates of the
astrophysical backgrounds. Any calculation that seeks to predict
these signals will be faced with uncertain theoretical inputs --
particularly the density of neutralinos at the extra-terrestrial
location of interest.

% ------------- Muons at IceCube ------------------
%
\begin{table}[t]
\begin{center}
\begin{tabular}{|c|c|} \hline
 Exposure [km$^2$ yr] & $\Phi_{10}$ [counts/(km$^2$ yr)] \\
\hline
0.2 & 50 \\
0.5 & 20 \\
1.5 & 6.7 \\
10 & 1.0 \\ \hline
\end{tabular}
\end{center}
{\caption{\label{table:muonfluxes}\footnotesize {\bf Muon Flux
Estimates for IceCube}. Flux of muons needed to produce 10 signal
events at IceCube assuming various exposures.}}
\end{table}
%---------------------------------------------------------

A good place to start, therefore, is at a location that is
well-understood such as at the center of the sun or earth. Relic
neutralinos can become gravitationally trapped in the core of the
sun, significantly enhancing their probability for annihilation. Of
the annihilation products, some fraction of the neutrinos can
eventually exit the sun and be detected in experiments such as
IceCube~\cite{Halzen:2006mq} via conversion of muon neutrinos into
muons.  To calculate this rate we integrate the differential flux of
conversion muons from solar-born as well as earth-born neutrinos
over the energy range $50\GeV \leq E_{\mu} \leq 300 \GeV$, assuming
an angular resolution of 3~degrees. The nominal target area for
IceCube is 1~km$^2$, but the effective area for detection of
neutrinos via muon conversion is smaller and can be cast as a
function of the muon
energy~\cite{GonzalezGarcia:2005xw,Halzen:2005ar,Barger:2007xf}. To
determine the prospects for observing muons at IceCube we estimate
various values of exposure in km$^2$-years needed for 10~signal
events, $\Phi_{10}$, as given in Table~\ref{table:muonfluxes}. We
note that the estimated background of muon events in one year of
IceCube data taking is expected to be
$\order(10)$~\cite{Barger:2007xf}.

%=(9)============ Parameter Set 4: Muons from neutrinos ===============
%\begin{comment}
\begin{figure}[t]
\begin{center}
\includegraphics[scale=0.55]{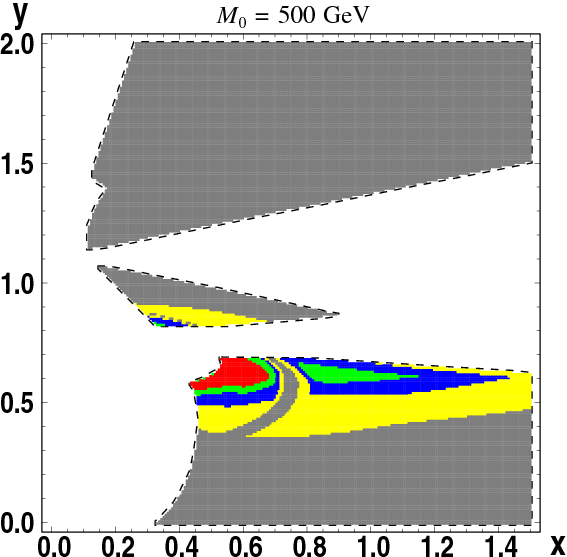}
\includegraphics[scale=0.46]{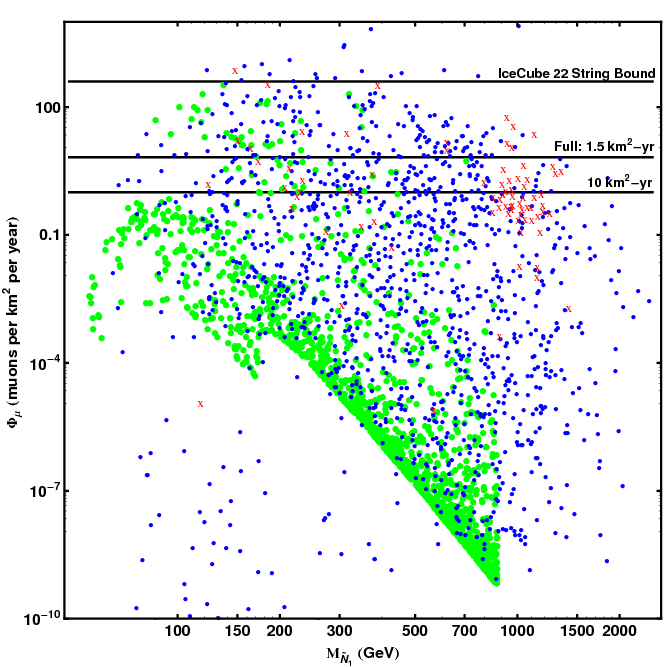}
\caption{\label{fig:muons}\footnotesize{\textbf{Upward Going Muon
Rates at IceCube}.} Left panel shows the IceCube reach for $M_0 =
500\GeV$. The red shaded region is detectable with 0.2~km$^2$-years
of exposure. Green and blue shaded regions would require 0.5 and
1.5~km$^2$-years of exposure, respectively. The yellow region gives
a visible signal after 10~km$^2$-years of exposure. The gray region
has a flux below 1~muon per km$^2$-yr and is likely undetectable at
IceCube. The right panel gives the muon flux at IceCube for the
random models defined by Table~\ref{tbl:scatter}. Points above the
horizontal line at $\Phi_{\mu}$ = 400 muons/km$^2$-year would likely
have given a signal in the 22-string IceCube data. The ultimate
reach for IceCube is approximated by the lowest horizontal line at
$\Phi_{\mu}$ = 1 muon/km$^2$-year.}
\end{center}
\end{figure}
%\end{comment}
%=================================================================

In Figure~\ref{fig:muons} we show the number of observed muons at
IceCube in units of counts per km$^2$-yr, combining those from
neutralino annihilation in the sun with those in the center of the
earth. The left panel shows the regions in the $\lbr x,\,y \rbr$
plane satisfying the condition $\Phi \geq \Phi_{10}$ for the values
listed in Table~\ref{table:muonfluxes} for the case when $M_0 = 500
\GeV$. For higher values of $M_0$ the (already low) muon rates drop
rapidly and little of the parameter space would be visible at
IceCube. This is evident in the right panel, which shows the muon
rate in units of muons/km$^2$-year for the random model set of
Table~\ref{tbl:scatter}. Recent results using a 22-string detector
at IceCube constrain the muon flux to be below 300-500 muons per
km$^2$-yr~\cite{Abbasi:2009uz}. None of the points in the left panel
of Figure~\ref{fig:muons} produce a flux this large, though a few of
the randomly-generated points would have produced a signal at
IceCube. We conservatively place the limit at 400 muons per
km$^2$-yr, indicated by the horizontal line in the right panel of
Figure~\ref{fig:muons}.

For low values of the mass scale $M_0$ there are some points which
can produce a flux of 50 muons per km$^2$-yr. These points are
indicated in the left panel Figure~\ref{fig:muons} by the red shaded
region. The three smaller flux values of Table~\ref{fig:muons} are
indicated by the green, blue and yellow regions, respectively. This
is roughly congruent with the parameter space that gives rise to a
mixed Higgsino/gaugino LSP and a light supersymmetric Higgs sector.
For the case where $M_0 = 1000\GeV$ and there is very little region
with a mixed LSP we expect the flux of muons at IceCube to be well
below 1~muon per km$^2$-yr throughout most of the parameter space.
The last two lines in Table~\ref{table:muonfluxes} are indicated by
the lower two horizontal lines in the right panel of
Figure~\ref{fig:muons}. Though the DMM models tend to give a larger
flux of muons at IceCube than unified models with an equivalent LSP
mass, the prospects for a visible signal at IceCube remain limited
for most of the parameter space.

%%%%%%%%%%%%%%%%%%%%%%%%%%%%%%%%%%%%%%%%%%%%%%%%%%%%%%%%%%%%%
%%%%%%%%%%%%%%% INDIRECT: GAMMAS %%%%%%%%%%%%%%%%%%%%%%%%%%%%
%%%%%%%%%%%%%%%%%%%%%%%%%%%%%%%%%%%%%%%%%%%%%%%%%%%%%%%%%%%%%
\subsection{Photons}
At distances further from the earth one can search for photons in
the gamma ray energy regime, which travel largely unimpeded from
their source. This allows gamma ray observatories to concentrate on
areas of the sky likely to have a high relic neutralino density --
such as the center of our galaxy. Photons can be produced as part of
decay chains or directly through loop-induced diagrams. The former
contribute to a continuous general spectrum of photons and are thus
more difficult to distinguish from astrophysical backgrounds. Direct
production of photon pairs -- or production of a single photon in
association with a Z-boson -- offers the possibility of a
monochromatic spectrum that can more easily be distinguished from
the background at the cost of a reduction in rate relative to the
continuous photon rate. We will consider searches for both phenomena
in this subsection.

The calculation of the flux of gamma rays observed from the
direction of the galactic center depends on the microscopic physics
of the neutralino and its interactions, but also very strongly on
the macroscopic physics of the halo profile assumed for the galaxy.
The latter is conveniently summarized by a single parameter $\bar
J\lp\Delta\Omega\rp$ where $\Delta\Omega$ represents the solid angle
resolution of the observatory. Here we will specifically consider
the Fermi/GLAST experiment for which we assume an angular resolution
of $\Delta\Omega=10^{-5}$~sr. We will consider the commonly adopted
profile of Navarro, Frank and White (NFW)~\cite{Navarro:1995iw}
which gives a value of $\bar J\lp \Delta\Omega \rp = 1.2644\times
10^4$, as well as a modified NFW profile that takes into account the
effect of baryons (a so-called ``adiabatic compression'' (AC)
model)~\cite{Blumenthal:1985qy,Mambrini:2005vk} for which we have
$\bar J\lp\Delta\Omega\rp=1.0237\times 10^6$. The large variation in
calculated rates for different halo profile models should always be
borne in mind when discussing ``predictions'' for gamma ray signals
of any particular model. Nevertheless, we can study the dependence
of the calculated rate on the underlying parameters and scale the
normalization of that rate to any particular halo profile via the
ratio of $\bar J\lp\Delta\Omega\rp$ values, as needed.

% ------------- Diffuse Photons Fermi ------------------
%
\begin{table}[t]
\begin{center}
\begin{tabular}{|c|c|c|} \hline
 Exposure [m$^2$ yr] & Halo Profile & $\Phi_{100}$ [counts/(cm$^2$ s)] \\
\hline
1 & NFW & $5.79\times10^{-10}$ \\
5 & NFW & $1.16\times10^{-10}$ \\
1 & NFW+AC & $7.15\times10^{-12}$ \\
5 & NFW+AC & $1.43\times10^{-12}$ \\ \hline
\end{tabular}
\end{center}
{\caption{\label{table:contgammafluxes}\footnotesize {\bf Reach
Estimates in the Continuous Gamma Ray Flux for the Fermi/GLAST
Experiment}. The quantity $\Phi_{100}$ is the flux needed for
100~signal events at Fermi/GLAST assuming various exposure times and
halo profiles.}}
\end{table}
%---------------------------------------------------------

For the diffuse signals we use {\tt DarkSUSY} to compute the
differential photon flux from neutralino annihilation in the halo in
units of photons/cm$^2$/s/GeV. We do so over the photon energy range
relevant for the Fermi/GLAST experiment of $1 \GeV \leq E_{\gamma}
\leq 200 \GeV$, with a step size of 1~GeV. Note that the upper limit
on the Fermi detector sensitivity is typically well below the mass
of the neutralino LSP for much of the parameter space we are
studying. We then numerically integrate these values to obtain an
overall photon flux in units of photons/cm$^2$/sec. Using the
effective area of 1~m$^2$ for the Fermi detector we compute the
minimum flux $\Phi_{100}$ necessary to produce an event rate of
100~signal photons from the galactic center in a given length of
exposure. As an example, using the NFW profile one obtains a value
of $\Phi_{100} = 5.79\times10^{-10}$ photons/cm$^2$/sec for one year
of running, where we assume data collection for 200~days per year.
If we require 100~signal photons over a five year mission this
becomes $\Phi_{100} = 1.16\times10^{-10}$ photons/cm$^2$/sec. To
take into account the possibility of a more generous halo profile
such as the NFW profile with adiabatic compression, we can simply
rescale this reach by the ratio of $\bar J\lp\Delta\Omega\rp$ for
the two profiles. In this case this implies a reach to lower fluxes
by a factor of approximately~80. The corresponding $\Phi_{100}$
values are listed in Table~\ref{table:contgammafluxes}.

%=(10)============ Parameter Set 4: Total photon flux ===============
%\begin{comment}
\begin{figure}[t]
\begin{center}
\includegraphics[scale=0.5]{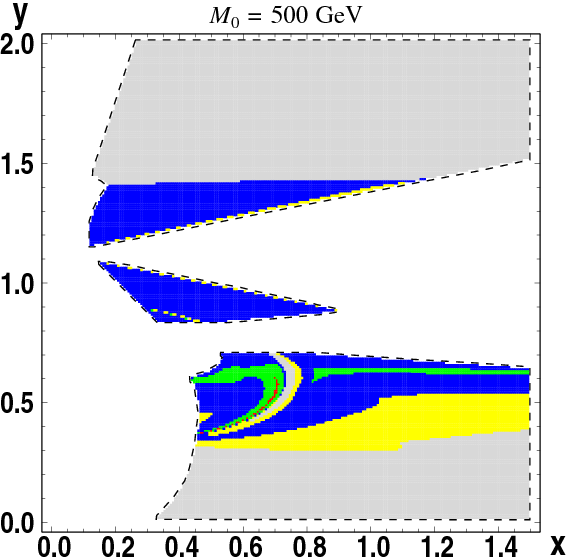}
\includegraphics[scale=0.5]{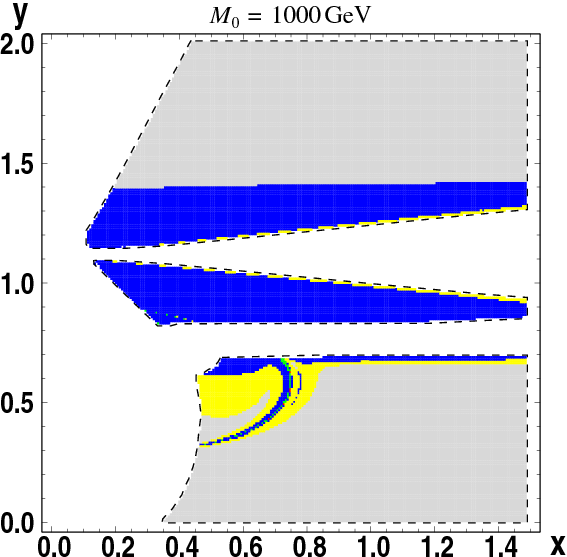}
\caption{\label{fig:gammas}\footnotesize{\textbf{Reach of the
Fermi/GLAST Experiment for Integrated Gamma Ray Flux}.} Left panel
is for $M_0 = 500\GeV$, right panel is for $M_0 = 1000 \GeV$.
Phenomenologically allowed areas are enclosed by the heavy dashed
lines. For $M_0=500 \GeV$ the green shaded region has $5.79 \times
10^{-10} > \Phi_{100} \geq 1.16 \times 10^{-10}$ photons/cm$^2$/sec
and therefore may be visible at Fermi/GLAST after five years of
exposure with the NFW profile. If the fluxes are rescaled to the
NFW+AC profile we obtain the reach shaded in blue (one year of
data-taking) and yellow (five years of data-taking). The gray shaded
area has an effective flux below $\Phi_{\gamma} = 0.014 \times
10^{-10}$ and is unlikely to yield a visible signal for diffuse
gamma rays at Fermi/GLAST.}
\end{center}
\end{figure}
%\end{comment}
%=================================================================

Translating this into a discovery reach for Fermi/GLAST requires
making an assumption about the background sources for gamma rays in
the $1 \GeV \leq E_{\gamma} \leq 200 \GeV$ energy regime. We will
use the estimate of~\cite{Bergstrom:1997fj} which gives a background
flux of $\Phi_{\rm bkgrnd} = 5.06\times 10^{-10}$
photons/cm$^2$/sec, or $\order(100)$ photons/year for the aperture
and energy range of Fermi/GLAST. This coincides with sensitivity
measures on the photon flux at Fermi/GLAST often quoted in the
literature~\cite{Morselli:2002nw,Hooper:2003ka}. In addition to
looking for excesses in the overall rate, the Fermi experiment will
also be able to use the shape of the observed spectrum to augment
signal extraction. Furthermore, there is reason to believe that
point sources near the galactic center can be identified and
subtracted from the signal to further enhance signal
significance~\cite{Dodelson:2007gd}. We will therefore
optimistically take the calculated flux $\Phi_{100}$ as a measure of
the reach of the Fermi experiment in the deflected mirage mediation
model space.

The result of this reach analysis is given in
Figure~\ref{fig:gammas} for the cases of $M_0 = 500\GeV$ (left
panel) and $M_0 = 1000\GeV$ (right panel). Detecting a gamma ray
signal in the deflected mirage mediation model will be challenging,
even with a relatively favorable set of assumptions about the halo
profile. For the low mass case of $M_0 = 500\GeV$ there is a small
region near the $A$-funnel (green shading) for which a signal is
possible after five years of data-taking at Fermi/GLAST with the NFW
profile. This coincides roughly with the region with a mixed
Higgsino/gaugino LSP. No such region exists in the higher mass $M_0
= 1000\GeV$ case. The blue and yellow regions in both panels are the
reach at Fermi/GLAST after one year and five years, respectively, if
the flux calculation was rescaled to the NFW + AC profile. Roughly
speaking, with this set of halo assumptions it should be possible to
detect a diffuse gamma ray signal over most of the parameter space
involving a wino-like or mixed Higgsino/gaugino LSP. The gray area
in the figures corresponds to the area with a bino-like LSP and
typically has an effective flux below $\Phi_{\gamma} = 0.014 \times
10^{-10}$ photons/cm$^2$/sec. These regions are unlikely to yield
visible signals without additional favorable assumptions about the
halo model. We note, however, that these rates would increase
considerably if we were to assume a local halo density normalized to
$\rho_0 = 0.3$ GeV/cm$^3$ regardless of the calculated thermal relic
density.

%=(11)============ Scatter: Photon Flux ========
%\begin{comment}
\begin{figure}[t]
\begin{center}
\includegraphics[scale=0.75]{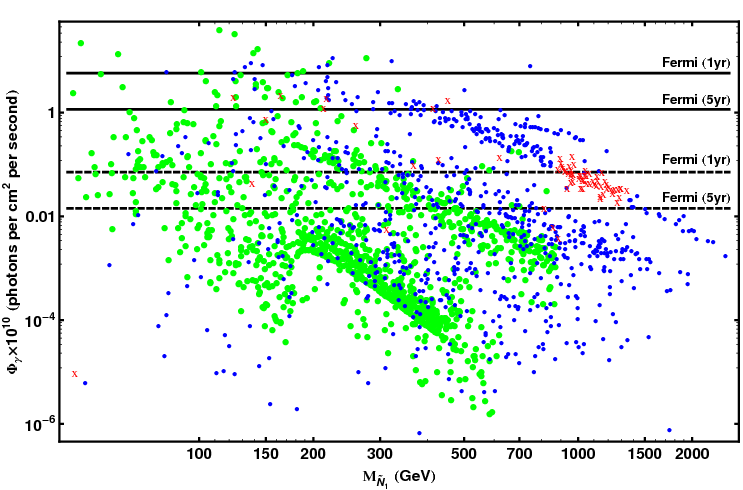}
\caption{\label{fig:gammas2}\footnotesize{\textbf{Integrated Gamma
Ray Flux for Models in Table~\ref{tbl:scatter}}.} Dark circles are
the ensemble of deflected mirage mediation models defined by
Table~\ref{tbl:scatter}. Larger lighter circles are an ensemble of
unified modulus-mediated models with $\alpha_m = \alpha_g = 0$. Red
crosses are the DMM models which fall into the ``WMAP preferred''
region~(\ref{prefer}). Solid lines indicate Fermi/GLAST reach
($\Phi_{100}$) after one year and five years of exposure. The dashed
lines indicate the same after rescaling all fluxes using the NFW +
AC profile.}
\end{center}
\end{figure}
%\end{comment}
%=================================================================

The best prospects for diffuse gamma ray signals occur near regions
that are ruled-out by certain scalars becoming the lightest
supersymmetric particle. This, in turn, is a property of the modular
weights assumed for the scalar sector of the theory. In
Figure~\ref{fig:gammas2} we show the results for the integrated
gamma ray flux in the range $1 \GeV \leq E_{\gamma} \leq 200 \GeV$
for the random collection of models defined by
Table~\ref{tbl:scatter} and modular weights~(\ref{nrandom}).
Choosing different modular weights does not produce a large number
of models with signals visible at Fermi/GLAST using the NFW profile.
The one-year and five-year $\Phi_{100}$ reach values are indicated
by the solid horizontal lines. More encouragingly, the models that
tended to satisfy the WMAP relic density constraints predicted
values of $\Phi_{100}$ that could potentially give observable
signals at Fermi/GLAST if the NFW + AC profile were applicable.

% -------------------------------------------------------
For the case of monochromatic signals, $\chi\chi\rightarrow
\gamma\gamma$ or $\chi\chi\rightarrow Z\gamma$, we obtain the flux
at a single energy, $E_\gamma=m_\chi$ or
$E_\gamma=m_\chi-\frac{m_z^2}{4m_\chi}$, respectively. As we have
seen previously, for the deflected mirage mediation scenario the
neutralino LSP can often be quite massive, particularly for the
cases that satisfy the WMAP thermal relic density constraint. These
monochromatic gamma ray signals are therefore likely to be at
energies beyond the reach of the orbiting Fermi/GLAST satellite
experiment. On the other hand, ground based atmospheric Cherenkov
telescopes (ACTs) such as CANGAROO~\cite{Cangaroo},
HESS~\cite{Hess}, MAGIC~\cite{Magic} and VERITAS~\cite{Veritas} have
thresholds for photon detection in the 100~GeV range and can detect
energetic photons up to $\order(10\TeV)$. Though data-taking is
restricted to dark, cloudless nights the effective area $A_{\rm
eff}$ of the target detector is quite large.

% ------------- Line Photons ACTs ------------------
%
\begin{table}[t]
\begin{center}
\begin{tabular}{|c|c|c|} \hline
 Exposure [m$^2$ yr] & Halo Profile & $\Phi_{10}$ [counts/(cm$^2$ s)] \\
\hline
1000 & NFW & $31.7\times10^{-15}$ \\
100 & NFW+AC & $3.92\times10^{-15}$ \\
500 & NFW+AC & $7.83\times10^{-16}$ \\
1000 & NFW+AC & $3.92\times10^{-16}$ \\ \hline
\end{tabular}
\end{center}
{\caption{\label{table:linegammafluxes}\footnotesize {\bf Reach
Estimates in the Monochromatic Gamma Ray Flux for a Generic ACT
Experiment}. Flux needed for 10 signal events at a generic ACT
assuming various exposures and halo profiles. Note that we are here
taking one year of data-taking to mean 365~days.}}
\end{table}
%---------------------------------------------------------

\begin{figure}[t]
\begin{center}
\includegraphics[scale=0.5]{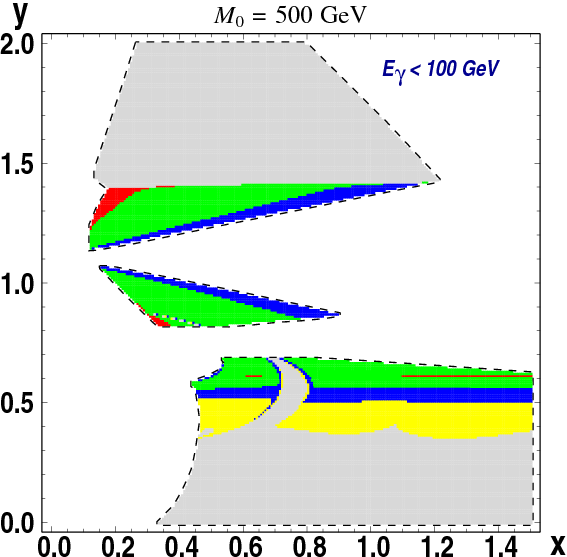}
\includegraphics[scale=0.5]{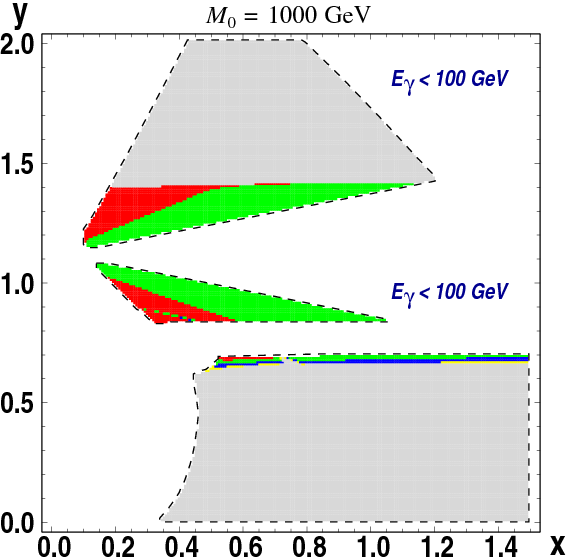}
\caption{\label{fig:mono}\footnotesize{\textbf{Reach of the
Fermi/GLAST Experiment for Monochromatic Gamma Ray Flux}.} Left
panel is for $M_0 = 500\GeV$, right panel is for $M_0 = 1000 \GeV$.
Red shaded regions near the edge of the allowed area in the two
plots have $\Phi_{10} \geq 31.7 \times 10^{-15}$ photons/cm$^2$/sec.
Rescaled to the NFW+AC profile we obtain the reach shaded in green
(100 m$^2$-yr exposure), blue (500 m$^2$-yr exposure) and yellow
(1000 m$^2$-yr exposure). Note that some of the indicated regions of
the parameter space give photon energies below the lower threshold
of $E_{\gamma} = 100\GeV$ and are therefore invisible to our generic
ACT. The gray shaded region has a monochromatic flux $\Phi_{10} \leq
0.39 \times 10^{-15}$ photons/cm$^2$/sec and is unlikely to give a
visible signal without a more favorable set of halo assumptions.}
\end{center}
\end{figure}

Unlike the case for continuum photons, the monochromatic signal has
virtually no background, particularly at high photon energies. We
therefore calculate the minimum flux $\Phi_{10}$ to produce 10
photons for a given exposure area $\times$ time, within the window
defined by the energy resolution of a typical ACT. We take this
energy resolution to be $\sigma_E/E = 0.15$. In
Figure~\ref{fig:mono} we present the reach for a generic ACT with
$A_{\rm eff} \simeq 3\times10^8$ cm$^2$ using the estimates for
$\Phi_{10}$ given in Table~\ref{table:linegammafluxes}.  Here we
have not included models with a neutralino LSP mass below 100 GeV,
as the ACTs do not typically take data below this energy.  As in
Table~\ref{table:contgammafluxes} we convert the quantity
$\Phi_{10}$ into an estimate of the reach of our generic ACT over
the model space for the NFW profile as well as for the NFW + AC
profile. Here we combine the photons from both the $\gamma\gamma$
and $\gamma\,Z$ signals as the 15\% energy resolution will not be
sufficient to resolve the two lines for LSP masses much in excess of
200~GeV.

%=(12)============ Parameter Set 4: Mono photon flux ===============
%\begin{comment}
%\end{comment}
%=================================================================

As in the case with the continuous gamma ray spectrum, detecting a
signal from neutralino annihilation will be difficult without help
from the halo profile. The red areas in the two plots have
$\Phi_{10} \geq 31.7 \times 10^{-15}$ photons/cm$^2$/sec and are
potentially visible with the NFW profile given 1000~m$^2$-years of
exposure. Note that the heavier mass cases ($M_0 = 1000\GeV$) are
easier to detect due to the sensitivity of ACTs to very energetic
photons. If the monochromatic fluxes are scaled to the values that
one would obtain if we had used the NFW + AC profile then the area
in green, blue and yellow shading (corresponding to $31.7 \times
10^{-15} > \Phi_{10} \geq 3.91 \times 10^{-15}$ photons/cm$^2$/sec,
$3.91 \times 10^{-15} > \Phi_{10} \geq 0.78 \times 10^{-15}$
photons/cm$^2$/sec and $0.78 \times 10^{-15} > \Phi_{10} \geq 0.39
\times 10^{-15}$ photons/cm$^2$/sec, respectively) would be visible
after 1000~m$^2$-years of exposure. This roughly corresponds to the
area of wino and mixed Higgsino/gaugino LSP.

%=(13)============ Scatter: Monochromatic ========
%\begin{comment}
\begin{figure}[t]
\begin{center}
\includegraphics[scale=0.75]{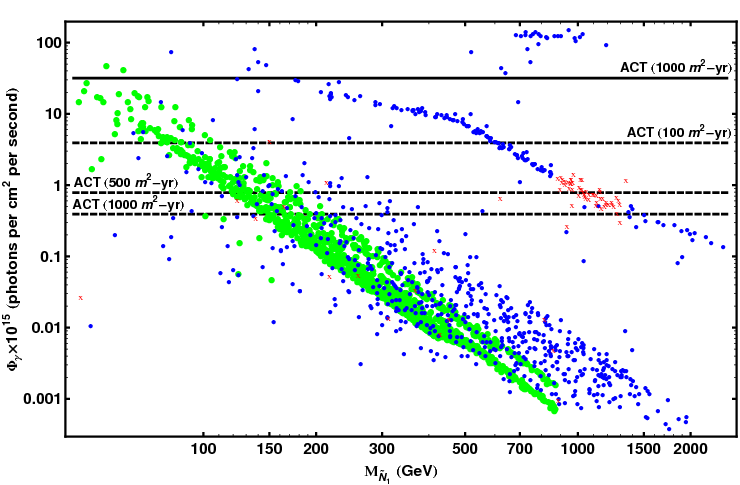}
\caption{\label{fig:mono2}\footnotesize{\textbf{Monochromatic Gamma
Ray Flux for Models in Table~\ref{tbl:scatter}}.} Dark circles are
the ensemble of deflected mirage mediation models defined by
Table~\ref{tbl:scatter}. Larger lighter circles are an ensemble of
unified modulus-mediated models with $\alpha_m = \alpha_g = 0$. Red
crosses are the DMM models which fall into the ``WMAP preferred''
region~(\ref{prefer}). The solid line indicates a generic ACT reach
($\Phi_{10}$) after 1000 m$^2$-years of exposure. The dashed lines
indicate the reach after rescaling all fluxes using the NFW + AC
profile for 100, 500 and 1000 m$^2$-years of exposure.}
\end{center}
\end{figure}
%\end{comment}
%=================================================================

In Figure~\ref{fig:mono2} we show the results for the monochromatic
gamma ray flux for the random collection of models defined by
Table~\ref{tbl:scatter} and modular weights~(\ref{nrandom}). The
reach $\Phi_{10}$ for a generic ACT after 1000 m$^2$-years of
exposure is given by the solid horizontal line. Dashed horizontal
lines give the reach after 100, 500 and 1000 m$^2$-years of
exposure, respectively, when all values of the flux are scaled up
using the NFW + AC profile. The rates of annihilation through loop
diagrams into mono-energetic photons are sensitive to the
wave-function of the LSP. Hence the deflected mirage mediation
models break into groups in Figure~\ref{fig:mono2} determined by the
makeup of the LSP. The unified models, on the other hand, are nearly
universal in having a bino-like LSP. We will return to this feature
in our discussion in Section~\ref{sec:compare}. As in
Figure~\ref{fig:gammas2} we see that the WMAP preferred models are
visible after a reasonable amount of ACT exposure provided the halo
profile has a sufficiently high concentration at the galactic
center.

%%%%%%%%%%%%%%%%%%%%%%%%%%%%%%%%%%%%%%%%%%%%%%%%%%%%%%%%%%%%%
%%%%%%%%%%%%%%% INDIRECT: ANTIMATTER %%%%%%%%%%%%%%%%%%%%%%%%
%%%%%%%%%%%%%%%%%%%%%%%%%%%%%%%%%%%%%%%%%%%%%%%%%%%%%%%%%%%%%
\subsection{Anti-matter in Cosmic Rays}

Dark matter annihilations will also produce charged particles, of
which positrons and anti-protons are the most promising for giving
signals above astrophysical backgrounds. There has been experimental
evidence for an excess of positrons in the 10-100 GeV range for some
time~\cite{Barwick:1995gv,Barwick:1997ig}. The recent data from the
PAMELA experiment seems to confirm the previous
results~\cite{Adriani:2008zr}. It is by now well appreciated that
standard MSSM models can fit the signal from PAMELA only with some
difficulty, and are even more challenged by the possible positron
excess at even higher energies reported by the ATIC
experiment~\cite{:2008zzr}.\footnote{As evidence for this statement
one can consider the analysis performed
in~\cite{Cirelli:2008pk,Donato:2008jk,Hooper:2008kv}, among many
other possible studies.} For that reason we will here only concern
ourselves with the positron and anti-proton energy ranges relevant
for PAMELA, roughly a few GeV up to 100~GeV.

The excess over background was reported at higher energies, so we
therefore consider the ten highest energy bins from the reported
signal. We will take as the central value of these bins the values
\begin{equation}
E_i = \lbr 5.55,\, 6.75,\, 8.25,\, 10.1,\, 13.05,\, 17.45,\,
23.85,\, 34.75,\, 53.2,\, 82.15\rbr \label{Ebins} \end{equation}
for the energies (in GeV) of the electron or positron cosmic ray.
Using {\tt DarkSUSY 5.0.4} (with all galactic propagation parameters
set to their default values) we calculate the differential positron
flux from neutralino annihilation with the NFW halo profile in units
of positrons/GeV/cm$^2$/sec/sr and obtain an average flux in each
bin, denoted $\psig$.  We then form the ratio of positron flux to
positron and electron flux as a function of the ``boost factor'' $B$
via the relation
\begin{equation} f_i(B)=\frac{\pbknd+B \psig}{\pbknd+B
\psig+\ebknda+\ebkndb}\, ,\end{equation}
where $\pbknd$ is the background positron flux and $\ebknda$,
$\ebkndb$ are the contributions to the electron flux from primary
and secondary sources, respectively.
Following~\cite{Moskalenko:1997gh,Baltz:1998xv} we use analytic
formulae to approximate these contributions from astrophysical
processes
\begin{equation} \pbknd=\frac{4.5 E^{0.7}}{1+650 E^{2.3}+1500E^{4.2}} \end{equation}
\begin{equation} \ebknda=\frac{0.16 E^{-1.1}}{1+11 E^{0.9}+3.2 E^{2.15}} \end{equation}
\begin{equation} \ebknda=\frac{0.7 E^{0.7}}{1+110 E^{1.5}+600E^{2.9}+580 E^{4.2}}~, \end{equation}
where $E$ represents the particle energy (normalized to units of
GeV).

To quantify how well a particular point in the parameter space of
the DMM model can fit the PAMELA data, we use the function $f(B)$
evaluated at each of the energy values in~(\ref{Ebins}). The
resulting quantities $f_i(B)$ are then compared to the observed
PAMELA data points $\fp_i$ using a chi-squared like function of the
boost factor
\begin{equation} \chi^2(B)=\sum_i
\left(\frac{\fp_i-f_i(B)}{\sigma_i}\right)^2~, \label{chiB}
\end{equation}
where the error values, $\sigma_i$, are taken to be the average
value of the full vertical error bar from the reported signal. We
then determine the value of the boost factor $B$ for which
$d\chi^2/dB=0$ and~(\ref{chiB}) is minimized.

%=(14)============ xy scan: Positrons ===============
\begin{figure}[p]
\begin{center}
\includegraphics[scale=0.52]{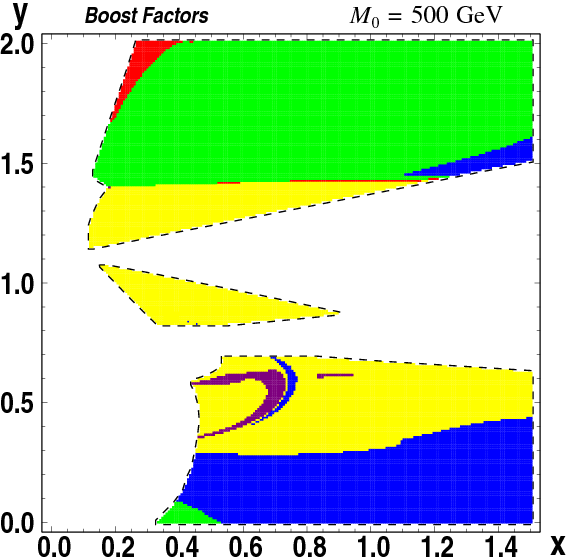}
\includegraphics[scale=0.52]{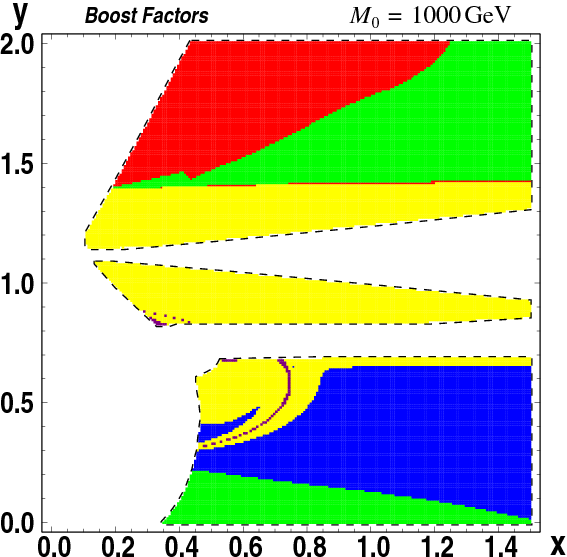}
\includegraphics[scale=0.52]{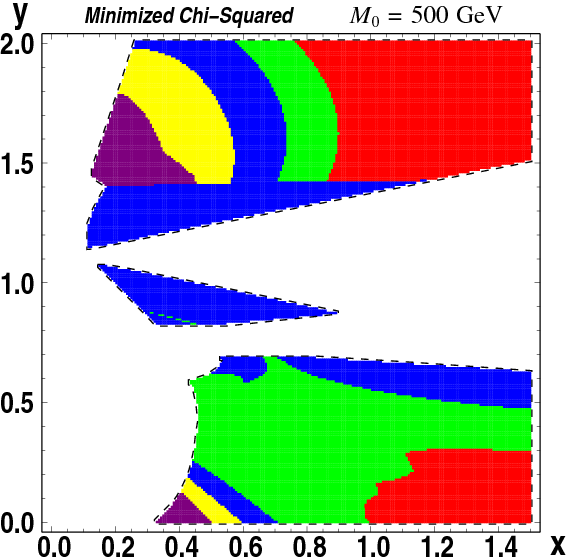}
\includegraphics[scale=0.52]{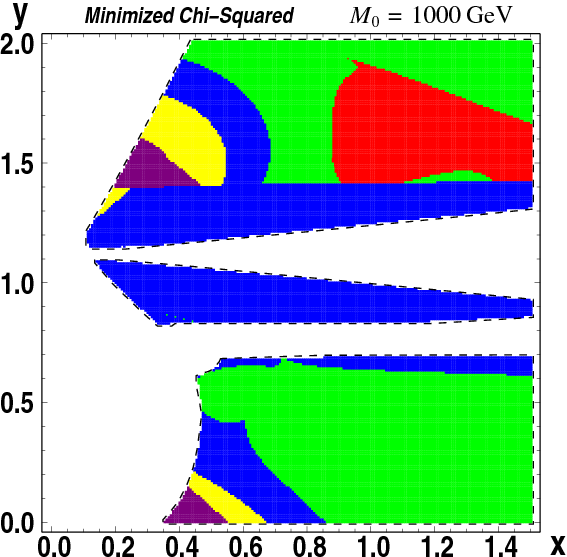}
\caption{\label{fig:positron}\footnotesize{\textbf{Best-Fit Boost
Factors and Minimized $\chi^2(B)$ Values for Fitting Anomalous
Positron Data}.} Left panels set $M_0 = 500\GeV$, right panels set
$M_0 = 1000 \GeV$. Phenomenologically allowed areas are enclosed by
the heavy dashed lines.
In the upper panels the regions are separated according to best-fit
boost values as follows: $B \geq 10^8$ red, $10^8 > B \geq 10^6$
green, $10^6 > B \geq 10^4$ blue, $10^4 > B \geq 10^2$ yellow, and
$B < 10^2$ in purple.
In the lower panels the regions are separated according to the
minimized $\chi_{\rm min}^2(B)$ values as follows: $\chi_{\rm
min}^2(B) \geq 2$ red, $2 > \chi_{\rm min}^2(B) \geq 1.5$ green,
$1.5 > \chi_{\rm min}^2(B) \geq 1$ blue, $1 > \chi_{\rm min}^2(B)
\geq 0.5$ yellow, and $\chi_{\rm min}^2(B) < 0.5$ in purple.}
\end{center}
\end{figure}
%\end{comment}
%=================================================================

%%%%%%%%%%%%%%%%%%%%%%%%%%%%
The results for the scan in the $\lbr x,\, y \rbr$ plane for
$M_0=500\GeV$ (left panels) and $M_0=1000\GeV$ (right panels) are
shown in Figure~\ref{fig:positron}. The values of the boost factors
that we obtain from minimizing $\chi^2(B)$ range from the relatively
innocuous $\order(10)$ to the preposterous $\order(10^9)$. The
largest boost factors occur for the heavy bino-like LSP in the red
shaded regions of the upper panels. The smallest boost factors are
denoted by the yellow and purple shaded regions and occur for the
wino-like LSP region (both panels) and the mixed Higgsino/neutralino
LSP region ($M_0 = 500\GeV$). Note that for these cases a large part
of what is here termed a ``boost factor'' is in fact simply the
result of a low thermal relic density. Invoking a non-thermal
mechanism for explaining the present cosmological density of such
wino-like LSPs would therefore result in $\order(1)$ boost factors
for many of the models in this part of the parameter space. This is
consistent with previous results suggesting that a low mass
wino-like LSP is the best candidate for an MSSM explanation for the
PAMELA data~\cite{Grajek:2008pg}.

The heavy bino-like LSP region fit the PAMELA data extremely well
(purple and yellow shaded regions in the lower panels of
Figure~\ref{fig:positron}), but this is merely an artefact of the
LSP being very massive -- the PAMELA signal is then interpreted as a
(massively-boosted) tail with a peak at a much higher energy level.
More realistic is the wino-like LSP region. Though the boost factors
were relatively small here, the calculated flux was not a
particularly good fit to the {\em shape} of the resulting positron
spectrum. Typical values of the minimized $\chi^2(B)$ were near
unity (the blue shaded region in the lower panels of
Figure~\ref{fig:positron}), but the $f_i(B)$ values tended to fall
with increasing energy like the background, as opposed to showing an
upward trend as seen by the PAMELA experiment. Nevertheless, when
certain reasonable modifications to the cosmic ray propagation
parameters in {\tt DarkSUSY} are made, the shape of the spectrum can
be made to fit the observations~\cite{Grajek:2008jb}.

The PAMELA experiment has also reported data on anti-protons in
cosmic rays which were consistent with expectations from background
astrophysical sources as well as previous experimental
data~\cite{Adriani:2008zq}. To compare the deflected mirage
mediation parameter space to this data we compute the differential
flux of anti-protons using {\tt DarkSUSY} with the NFW profile and
default diffusion parameters for the ten highest bins reported
in~\cite{Adriani:2008zq}, corresponding to anti-proton mean kinetic
energies of 5.85, 6.98, 8.37, 10.1, 12.3, 15.3, 19.5, 25.9, 37.3 and
61.2~GeV. We use the estimations of Cirelli et
al.~\cite{Cirelli:2008id} for the background flux from astrophysical
processes. We consider a model point to be in conflict with the
PAMELA data if the predicted signal flux in any one of the ten bins
were twice as high as the reported experimental observation. None of
the parameter space gives rise to an anti-proton flux of this
magnitude. If, on the other hand, we were to naively apply the same
boost factor to the anti-proton flux as we do to the positron flux
then the majority of these models would be in conflict with the
PAMELA anti-proton data. While anti-protons and positrons will
propagate through the galaxy differently~\cite{Maurin:2002uc}, it is
reasonable to assume that the fluxes will involve similar
astrophysical backgrounds. Thus the two sets of data appear to be in
conflict throughout the DMM parameter space unless modifications to
the diffusion model from the {\tt DarkSUSY} default values are made.

%%%%%%%%%%%%%%%%%%%%%%%%%%%%%%%%%%%%%%%%%%%%%%%%%%%%%%%%%%%%%
%%%%%%%%%%%%%%% BENCHMARKS %%%%%%%%%%%%%%%%%%%%%%%%%%%%%%%%%%
%%%%%%%%%%%%%%%%%%%%%%%%%%%%%%%%%%%%%%%%%%%%%%%%%%%%%%%%%%%%%
\subsection{Benchmarks}

Before concluding this section we will look back at the dark matter
signatures for the specific benchmarks outlined in
Table~\ref{tbl:models} in Section~\ref{sec:space}. To remind the
reader, models~A, B~and~C are specific examples presented in the
original papers on deflected mirage
mediation~\cite{Everett:2008qy,Everett:2008ey}. Models D~and~E are
chosen from the results of our parameter space survey, while model~F
is a point without gauge-charged messengers with $\alpha_m = 1$ as
in the KKLT model. The properties of the neutralino LSP, the masses
of other key particles, and the values for the thermal relic density
and other dark matter observables are given in
Table~\ref{table:benchmarks2}.

% ------------- benchmark properties v2 units in caption ------------------
%
\begin{table}[t]
\begin{center}
{\footnotesize \begin{tabular}{|c||c|c|c|c|c|c|}\hline
  & Model A & Model B & Model C & Model D & Model E & Model F \\ \hline\hline
 $m_{\LSP}$ & 1009 & 626.6 & 772.8 & 257.2 & 193.2 & 710.8 \\
 $f_B$ & 0.912 & 0.989 & 0.983 & 0.914 & 0.000 & 0.932 \\
 $f_W$ & 0.071 & 0.001 & 0.001 & 0.030 & 0.997 & 0.004 \\
 $f_H$ & 0.017 & 0.009 & 0.017 & 0.057 & 0.003 & 0.064 \\ \hline
 NLSP & $\wt{C}_1$ & $\wt{C}_1$ & $\wt{\tau}_1$ & $\wt{C}_1$ & $\wt{C}_1$ & $\wt{\tau}_1$ \\
 $m_{\rm NLSP}$ & 1026 & 743.8 & 781.3 & 302.6 & 193.3 & 762.8 \\
 $m_h$ & 119.5 & 120.6 & 121.5 & 112.1 & 119.8 & 119.9 \\
 $m_A$ & 1559 & 1043 & 1136 & 473.5 & 1518 & 1136 \\ \hline
 $\Ochi$ & 0.117 & 0.891 & 0.517 & 0.088 & 0.001 & 0.537 \\
 $R_{\rm Xe}$ & 0.046 & 0.022 & 0.032 & 1.182 & $7\times 10^{-5}$ & 0.178 \\
 $R_{\rm Ge}$ & 0.037 & 0.018 & 0.025 & 0.881 & $6\times 10^{-5}$ & 0.140 \\ \hline
 $\Phi_{\gamma\gamma}$ & $1.778\times 10^{-17}$ & $3.517\times 10^{-17}$ & $2.734\times 10^{-17}$ & $4.128\times 10^{-16}$ & $2.071\times 10^{-15}$ & $2.267\times 10^{-17}$ \\
 $\Phi_{\gamma Z}$ & $9.169\times 10^{-17}$ & $3.244\times 10^{-18}$ & $3.326\times 10^{-18}$ & $8.868\times 10^{-16}$ & $6.399\times 10^{-15}$ & $1.157\times 10^{-17}$ \\
 $\Phi_{\rm tot}$ & $1.846\times 10^{-12}$ & $2.260\times 10^{-12}$ & $2.041\times 10^{-12}$ & $1.449\times 10^{-10}$ & $1.463\times 10^{-11}$ & $4.319\times 10^{-12}$ \\ \hline
 $\Phi_\mu$  & 0.738 & 0.194 & 0.495 & 11.766 & 0.016 & 3.616 \\
 $\Phi_{\bar D}$  & $8.096\times 10^{-15}$ & $2.155\times 10^{-14}$ & $1.195\times 10^{-14}$ & $1.631\times 10^{-12}$ & $2.824\times 10^{-14}$ & $2.560\times 10^{-14}$ \\
 $\Phi_{\bar p}$  & $7.438\times 10^{-11}$ & $9.979\times 10^{-11}$ & $8.923\times 10^{-11}$ & $6.523\times 10^{-9}$ & $8.617\times 10^{-10}$ & $1.887\times 10^{-10}$ \\
 $B$ & 4774 & 5156 & 5145 & 102.9 & 1147 & 2510 \\
%
% $\chi^2|_B$ & 1.360 & 1.564 & 1.512 & 1.797 & 1.119 & 1.548 \\
 \hline
\end{tabular}}
\end{center}
{\caption{\label{table:benchmarks2}\footnotesize {\bf
Characteristics of Benchmark Models From Table~\ref{tbl:models}}.
All masses are given in GeV and we denote wave-function components
as $f_B=|N_{11}|^2$, $f_W=|N_{12}|^2$ and
$f_H=|N_{13}|^2+|N_{14}|^2$. Direct detection rates $R_i$ have units
of recoils/kg-year. Photon fluxes are given in units of
photons/cm$^2$/sec and $\Phi_{\rm tot}$ represents the diffuse gamma
flux integrated from~1 to 200~GeV. The muon flux is given in units
of muons/km$^2$-year. $\Phi_{\bar D}$ is given in units of
anti-deuterons/(GeV s cm$^2$ sr) and is computed at 0.25~GeV.
$\Phi_{\bar p}$ is given in units of anti-protons/(GeV s cm$^2$ sr)
and is computed at 10~GeV.}}
\end{table}
%---------------------------------------------------------

Properties of the LSP are given in the first block of entries in
Table~\ref{table:benchmarks2}. The wave-function of the LSP is
parameterized as
\begin{equation}
\wtd{N}_{1} = N_{11} \tilde{B} + N_{12} \tilde{W} + N_{13}
\tilde{H}^{0}_{d} + N_{14} \tilde{H}^{0}_{u}, \label{LSPcontent}
\end{equation}
which is normalized to $N_{11}^2+N_{12}^2+N_{13}^2+N_{14}^2=1$. The
properties of the NLSP and key Higgs masses are given in the next
block of entries. These can have a significant impact on the thermal
relic density calculation, which is provided in the next group of
entries. All of the benchmark models have a thermal relic density
that is below unity, though only models~A and~D agree well with the
WMAP results of~(\ref{omegah2}). In computing the relic density
values all the models receive some reduction in abundance due to
coannihilation. This is particularly true for points~A, C~and~E. For
this last case the LSP is mostly wino-like and the computed relic
density is well below the WMAP bound. It is also below the threshold
value of $\Ochi_{\rm min} = 0.025$ and thus all subsequent dark
matter signals have been rescaled to take this into account. Model~D
is an example of a point which resides in the $A$-funnel region
where $2m_{\LSP}\simeq m_A$ which also has effects on other dark
matter observables.

Apart from the wino-like model~E, all of these cases would be probed
by direct detection experiments at the one-ton level, and most would
give a sizable signal for detectors in the 100-300 kilogram range.
As previously mentioned, one should keep in mind that these rate
values do come with large theoretical uncertainties associated with
the nuclear matrix element factors which go into the calculation.
Models~D and~F have the largest Higgsino content and hence the
largest direct detection rates, with model~D giving the most sizable
rate due to its low mass LSP. We point out that the very low rate
for model~E is in part due to the rescaling of the local halo
density we performed. If we had simply assumed a local halo density
normalized to 0.3~GeV/cm$^3$ these rates would be increased by a
factor of~25.

Gamma ray signals for the benchmarks are shown using the NFW halo
profile. For the continuous gamma ray signal we estimate that a flux
on the order of 1-5$\times10^{-10}$ photons/cm$^2$/sec in the
Fermi/GLAST photon energy range of~1 to~200 GeV is needed to observe
a signal over the astrophysical background. Only model~D gives rise
to such a large flux of gamma rays. Each of the others, however,
would be observable above background if the NFW profile with
adiabatic compression were employed. The wino-like LSP of model~E
gives a large signal for the two monochromatic gamma ray signals.
Such a (combined) signal is just above the threshold for most ACT
experiments and would give a visible signal with the NFW profile for
5000 m$^2$-years of exposure -- not an unreasonable amount given the
size of most atmospheric Cherenkov detectors. For the NFW + AC
profile it is likely that only models~A, D~and~E will give rise to a
detectable monochromatic gamma ray signal. Expectations for
observation of these benchmarks at IceCube via the flux of muons
from neutralino annihilation in the sun are also a bit mixed.
Models~D and~F have the highest muon fluxes and ought to produce
muons at rates above background for reasonable estimates of the
effective area of the IceCube detector at these energies. The
remaining models are unlikely to produce a signal even after
10~km$^2$-years of data-taking at IceCube.

Anti-matter prospects fare slightly worse across our benchmark
models. In all cases the naive estimate of anti-proton fluxes is
well below the background expectation of
$\order(10^{-7})\,\bar{p}$/(GeV s cm$^2$ sr) at our reference point
of 10~GeV of kinetic energy for the anti-proton. If one wishes to
explain the PAMELA positron data using these models, however, one is
confronted with rather large boost factors for all models save
model~D. Naively applying such boost factors to the anti-proton flux
would place all of these models in tension with the anti-proton
data, giving a signal flux of roughly an order of magnitude above
the background. For completeness we have also included the flux of
anti-deuterons computed at a reference point of 0.25~GeV for the
$\bar{D}$ kinetic energy which is relevant for the planned GAPS
experiment~\cite{Hailey:2005yx}. Only model~D again is likely to
give a sizable signal in this indirect detection channel.

\noindent\section{Mirage Mediation versus Deflection Mirage
Mediation}
\label{sec:compare}

In the previous section we looked at how a number of
astroparticle/cosmological signals depend on the parameters of the
deflected mirage mediation paradigm. We did this in terms of the
theory motivated $\lbr x,\,y \rbr$ parameter space where we saw that
the detection prospects are good for general values of the parameter
$x$ when $0.4 \lappeq y \lappeq 1.4$ for the modular weight
choice~(\ref{weights}). We also averaged over the other more
model-dependent parameters in the theory, presenting our results in
the form of scatter plots, using a unified mSUGRA-like theory as a
foil with which to compare the predictions of the DMM model. That
DMM and unified models give different ``footprints'' should not be a
surprise, since the importance of non-universality in gaugino masses
for dark matter observables has been emphasized for some
time~\cite{BirkedalHansen:2001is,BirkedalHansen:2002am}. More
interesting is the question of whether observations in the dark
matter arena are capable of distinguished DMM from the
(un-deflected) mirage pattern which preceded it.

\begin{figure}[t]
\begin{center}
\includegraphics[scale=0.58]{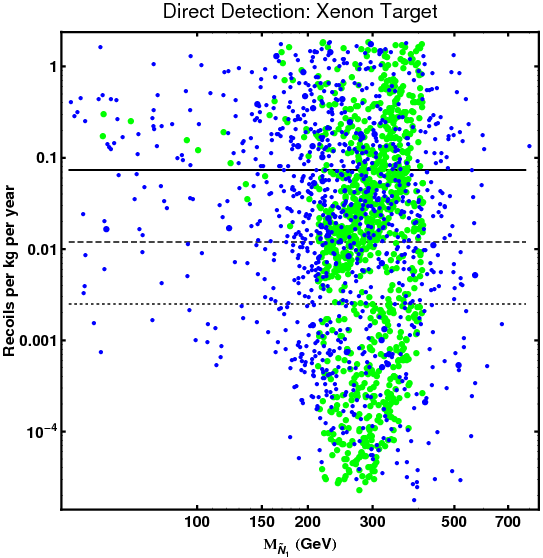}
\includegraphics[scale=0.58]{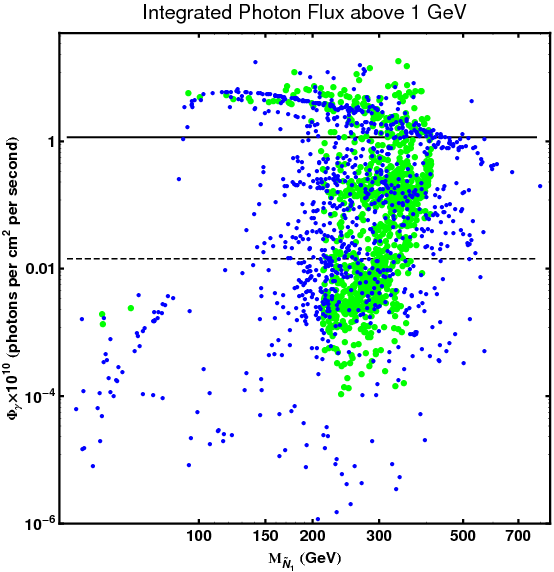}
\caption{\label{fig:MvsD}\footnotesize{\textbf{Comparison of
Deflected Mirage Mediation with the Case of No Messengers}.} A
random sample of models based on the ranges of
Table~\ref{tbl:scatter} with $M_0$ fixed at 500 GeV for the
deflected mirage case (dark points) versus the same ranges but with
$N_m = 0$ (light points). The left plot gives the recoil rate per
kg-year on xenon targets with sensitivities for Xenon100 after one
year (solid line), LUX after three years (dashed line), and Xenon1T
after five years (dotted line). The right figure gives the photon
flux for all energies above 1~GeV from the galactic center. The
solid line is the reach of Fermi/GLAST after five years with the NFW
profile while the dashed line is the  five year reach with the NFW +
AC profile.}
\end{center}
\end{figure}

Looking back at the figures from the previous section we might
conclude that the answer is certainly affirmative. After all, when
we move away from the mirage prediction $x+y=1$ we see strong
variation in both the mass and composition of the LSP, resulting in
dramatic changes in the thermal relic density of the neutralino.
Moving away from the $x+y=1$ limit can increase recoil rates at
direct detection experiments and change the strength of signals at
indirect detection experiments. But this is somewhat deceptive,
since in those figures a number of important parameters --
particularly the overall scale $M_0$ -- were held constant. When we
average over these values the distinction is far less clear. To
exemplify this, consider Figure~\ref{fig:MvsD} in which we look at
two of the more commonly considered observables in the literature:
recoil rates on xenon targets and the integrated photon flux above
1~GeV in energy from the direction of the galactic center. In this
figure we have fixed the value of $M_0 = 500\GeV$. The dark points
continue to represent a random collection of~1000 deflected mirage
mediation models with parameters taken from the ranges given in
Table~\ref{tbl:scatter}. The lighter shaded points are now no longer
a unified model, but instead represent an additional 1000~points
with the same ranges of Table~\ref{tbl:scatter} except that we fix
$N_m = 0$. These are, therefore, mirage models with a similar
overall mass scale. Though the range in LSP mass is more restricted
in the un-deflected case, the signature range expected is nearly
identical. These examples could be replaced with many of the signals
we considered in Section~\ref{sec:signals}. Such ``inclusive'' style
measurements in the cosmological arena will not in themselves be
able to determine the presence of the gauge mediation component in
the underlying theory.

%=(15)============ Parameter Set 4: Anti-deuteron rates ========
%\begin{comment}
%\end{comment}
%=================================================================

The reason for this degeneracy is readily found by considering
another way to parameterize the gaugino masses in the DMM model. We
can translate each of the instances of the running gauge coupling
$g_a^2(\mu)$ in~(\ref{Mafull}) to the high scale $\mu_{\UV} =
\mu_{\GUT}$ via the relation
\begin{equation} g_a^2\(\mu_{\UV}\) = \frac{g_a^2(\mu)}{1-\frac{g_a^2(\mu)}{8\pi^2}\ln\lb
\lp \frac{\mu_{\GUT}}{\mu}\rp^{b_a}\lp\frac{\mu_{\GUT}}{\mu_{\rm
mess}}\rp^{N_m} \rb} \, . \end{equation}
The gaugino masses can now be written
\begin{equation} \frac{M_a(\mu_{\EW})}{g_a^2(\mu_{\EW})} =
\frac{M_0}{g_a^2\(\mu_{\UV}\)}\[1 + \alpha_m
\frac{g_a^2\(\mu_{\UV}\)}{16\pi^2}\(b_a - \alpha_g
N_m\)\lnf{\mpl}{\mgrav}\] \label{Ma1a} \, ,\end{equation}
where the first term in the square brackets is precisely the case of
gaugino mass unification at the high scale, as in the minimal
supergravity paradigm. As in~\cite{Choi:2007ka,Altunkaynak:2009tg}
we see that departures from universality are governed by the
parameter $\alpha_m$, but the slopes of these departure trajectories
are altered relative to the case of simple mirage mediation. We can
make this more explicit by defining new variables
\begin{equation} \wtd{\alpha} \equiv \[\alpha_m
\frac{g_a^2\(\mu_{\UV}\)}{16\pi^2}\lnf{\mpl}{\mgrav}\]^{-1} - N_m
\alpha_g\, ; \quad \quad \wtd{M}_0 \equiv \alpha_m
\frac{M_0}{16\pi^2}\lnf{\mpl}{\mgrav} = \frac{\mgrav}{16\pi^2}
\label{tildedef}
\end{equation}
such that~(\ref{Ma1a}) becomes
\begin{equation}
\frac{M_a(\mu_{\EW})}{g_a^2(\mu_{\EW})} = \wtd{M}_0\(\wtd{\alpha} +
b_a\) \label{Ma1b} \, .\end{equation}
From~(\ref{Ma1b}) it is immediately apparent that despite the added
complexity of the gauge-charged messenger sector the gaugino masses
continue to depend on only a single dimensionless parameter in
addition to an overall mass scale, here given by
$\wtd{M}_0$.\footnote{Note that this variable is equivalent to $M_s$
defined in~\cite{Falkowski:2005ck} and utilized
by~\cite{Baer:2006id,Baer:2007eh}.} In other words, for every mirage
model with $N_m = 0$ there exists a family of deflected mirage
models with {\em the same set} of soft supersymmetry breaking
gaugino masses at the electroweak scale. Any differences in
phenomenology will therefore be due to the differences in scalar
masses and the resultant EWSB parameters.

%---------------- Look-alike Table --------------------
\begin{table}[t]
\begin{center}
\begin{tabular}{|c||c|c||c|c|c|} \multicolumn{6}{c}{Parameters} \\ \hline
 & $\wtd{\alpha}$ & $\wtd{M}_0$ [GeV] & $M_0$ [GeV]
 & $\alpha_m$ & $N_m \times \alpha_g$ \\
\hline
Mirage & 6.52 & 587 & 1976 & 1.5 & 0  \\
DMM & 6.52 & 599 & 889 & 3.4 & -4.3 \\
\hline \multicolumn{6}{c}{Physical Quantities} \\ \hline
 & H\% & $M_{\chi}$ [GeV] & $\Ochi$ & $\sigma_{\rm SI}^{p} [{\rm cm}^2] $
 & $\Phi_{\gamma}^{\geq 1\GeV} [{\rm cm}^{-2}{\rm s}^{-1}]$ \\
 \hline
Mirage & 98.1\% & 1247 & 0.13 & $0.63\times 10^{-45}$ &
$6.51\times 10^{-12}$ \\
DMM & 99.5\% & 1040  & 0.10 & $2.86\times 10^{-45}$ &
$2.41\times 10^{-12}$ \\
\hline
\end{tabular}
\end{center}
{\caption{\label{tbl:degen}\footnotesize {\bf Two Degenerate Models
from the Mirage Family}. The mirage model and the DMM model have
nearly identical values for the gaugino mass-determining parameters
of equation~(\ref{tildedef}). Both have nearly Higgsino-like LSPs of
approximately a TeV in mass. Both give rise to a thermal relic
abundance of LSP neutralinos in line with the WMAP preferred region
of~(\ref{prefer}) and have similar predictions for the observables
in Figure~\ref{fig:MvsD}.}}
\end{table}
%------------------------- END OF THE TABLE ---------------------

The parametrization in~(\ref{Ma1b}) makes it easy to construct such
look-alike ``degenerate'' points. Consider the two example points in
Table~\ref{tbl:degen}. The (un-deflected) mirage model and the DMM
model have nearly identical values for the key parameters
$\wtd{\alpha}$ and $\wtd{M}_0$ of~(\ref{tildedef}) despite having
very different underlying physics inputs. The mirage model has
modular weights $n=1/2$ for all fields and $\tan\beta = 2.5$ while
the deflected mirage model has vanishing modular weights and
$\tan\beta = 27.8$. Both are in accordance with the WMAP
``preferred'' relic abundance in equation~(\ref{prefer}) and have
similar predictions for the two observables plotted in
Figure~\ref{fig:MvsD}. In particular, both would give an observable
signal in one ton-year of exposure on a xenon target but neither
would be detectable above background at Fermi/GLAST with the NFW
halo profile (though both would be very nearly detectable if the NFW
+ AC profile was used). Their predictions for other observables are
also similar in magnitude.

Though many such degenerate pairs can be constructed, it is not the
case that there are {\em no} signals associated with relic
neutralinos that are sensitive to the presence of gauge mediation in
the underlying model. We will here merely point out two such
examples. To do so we will change variables once again to a
dimensionless quantity that is directly proportional to the number
of gauge-charged messengers in the theory. Writing out~(\ref{MaC})
in terms of the explicit mass scales in the theory as
\begin{equation}
M_a(\mu)=M_0\lc 1+\beta_a(\mu)\lb t-\frac{N_m}{b_a}
\lnf{\mu_{\GUT}}{\mu_{\rm mess}} \rb\rc -
\frac{\beta_a(\mu)N_m}{2b_a}\lmess +\frac{\beta_a(\mu)}{2}\mgrav \,
, \label{general_low}
\end{equation}
we may apply the definitions~(\ref{alphag}) and~(\ref{alpham}) to
write~(\ref{general_low}) as
\begin{equation}
M_a(\mu) =M_0\lb 1+\beta_a(\mu)t \rb
+\mgrav\frac{\beta_a(\mu)}{2}\lb 1-\frac{\alpha'}{b_a} \rb\, ,
\label{Ma3} \end{equation}
where we have introduced a new dimensionless parameter $\alpha'$
defined as
\begin{equation}
\alpha' = N_m \lb \ag+\frac{2}{\am}\frac{\ln\lp \mu_{\GUT}/\mu_{\rm
mess} \rp}{\ln\lp M_{\PL}/\mgrav \rp} \rb\, . \label{alphap}
\end{equation}
%

%=(16)============ Compare: Photon Energies ========
%\begin{comment}
\begin{figure}[t]
\begin{center}
\includegraphics[scale=0.75]{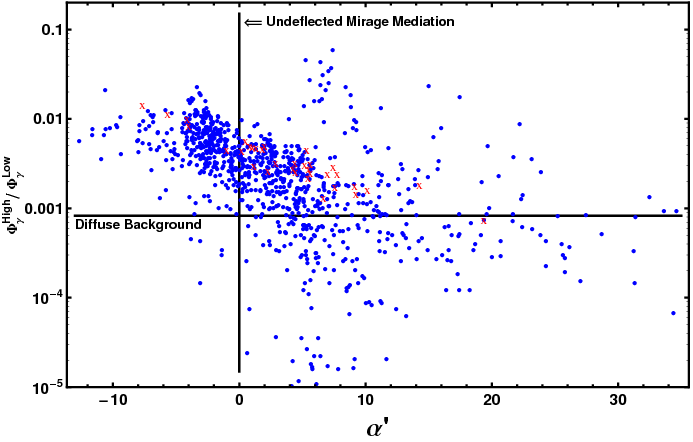}
\caption{\label{fig:compare1}\footnotesize{\textbf{Hardness of the
Gamma Ray Spectrum from the Galactic Center as a Function of
$\alpha'$}.} Flux of photons in the energy range $60\GeV \leq
E_{\gamma} \leq 200\GeV$ divided by the flux in the range $1\GeV
\leq E_{\gamma} < 60\GeV$ is plotted for the signal photons versus
the parameter $\alpha'$ of equation~(\ref{alphap}). The same ratio
for the background is given by the heavy horizontal line. Note the
logarithmic scale of the vertical axis.}
\end{center}
\end{figure}
%\end{comment}
%=================================================================

Although specifying the value of $\alpha'$ does not uniquely
determine the underlying model, all mirage models will have
$\alpha'=0$, which makes this a convenient parametrization for our
purpose here. When one plots inclusive counting signatures as a
function of $\alpha'$ -- such as the number of recoils per kg-year
on xenon targets or the total integrated flux of photons coming from
the galactic center -- the results are much as in
Figure~\ref{fig:MvsD}. But certain exclusive measurements reveal
some correlation. For example, one can take advantage of the
Fermi/GLAST experiment to measure the spectral profile of detected
gamma rays. The shape of this spectrum carries information about the
mass and wave-function content of the annihilating neutralino LSP.
We can attempt to take this information into account by using {\tt
DarkSUSY} to compute the differential photon flux in 1~GeV
increments over the energy range $1\GeV \leq E_{\gamma} \leq
200\GeV$. From this information we create an interpolating function
which is then integrated over six energy bins: 1~-~10~GeV,
10~-~30~GeV, 30~-~60~GeV, 60~-~100~GeV, 100~-~150~GeV, and
150~-~200~GeV.

In Figure~\ref{fig:compare1} we take the ratio of the total flux
over the three highest energy bins to the total flux over the three
lowest energy bins and plot it as a function of the parameter
$\alpha'$. Over the range $-10 \leq \alpha' \leq 10$ (which is the
regime of moderate $\alpha_g$ values) the hardness of the photon
spectrum is reasonably well correlated with the value of $\alpha'
\simeq N_m \times \alpha_g$. Note that over this range of parameters
we generally expect the gamma rays from neutralino annihilation to
have a harder spectrum than those coming from background
astrophysics sources, as indicated by the heavy horizontal line. The
collection of models in Figure~\ref{fig:compare1} was generated
using a fixed mass scale of $M_0 = 500\GeV$. The degree of
correlation increases if we choose higher values for this parameter.

%=(17)============ Compare: Mono Lines ========
%\begin{comment}
\begin{figure}[t]
\begin{center}
\includegraphics[scale=0.75]{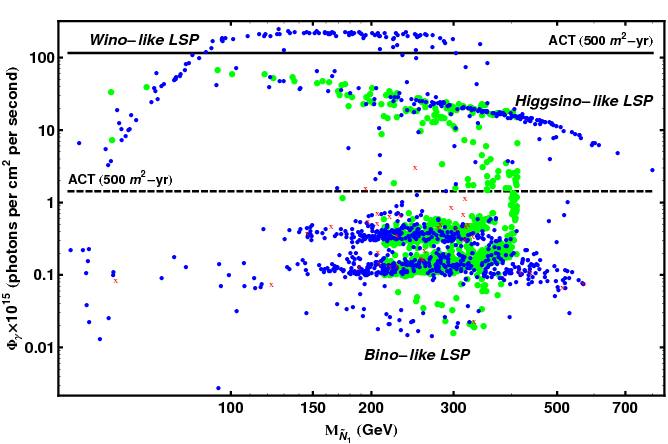}
\caption{\label{fig:compare2}\footnotesize{\textbf{Combined
Monochromatic Line Flux as a Function of Neutralino Mass}.} The
solid line represents the reach at a generic ACT for 500 m$^2$-years
of exposure with an NFW profile. The dashed line is the reach for
the same exposure assuming an NFW + AC profile.}
\end{center}
\end{figure}
%\end{comment}
%=================================================================

More generally, the presence of the gauge-charged messengers allows
the theory space to explore neutralino LSPs with differing
wave-function composition from that of the mirage case with $N_m =
0$. This is already apparent from Figure~\ref{fig:mono2}. Such
effects are known to have a large impact in the rate for
loop-induced annihilation into monochromatic photons. In
Figure~\ref{fig:compare2} we give the combined flux at both the
$\gamma\gamma$ and $\gamma\,Z$ energies for our random sample of DMM
models with $M_0 = 500 \GeV$ (dark points) and our random sample of
mirage model (light points) at the same $M_0$ value. When the
overall mass scale $M_0$ is held fixed the points tend to cluster as
a function of the neutralino wave-function composition. In the
mirage mediation model without gauge messengers the majority of
models have a bino-like LSP with a smaller grouping of Higgsino-like
LSPs. In fact, the maximal value of the wino-component of the LSP
was typically 20\%, and these occurred mostly in cases with
vanishing modular weights for the matter multiplets and $n=1$ for
the Higgs multiplets. By contrast the DMM models are able to achieve
a wino-like LSP across all modular weight choices. Assuming an NFW
halo profile only those cases where the LSP is nearly 100\%
wino-like have a chance to be observed in a typical ACT experiment.
Assuming an adiabatically-compressed profile allows most of the
wino-like and Higgsino-like cases to be detected in 500 m$^2$-years
of exposure.

\section{Conclusion}

Deflected mirage mediation, like the simpler mirage pattern which
preceded it, are motivated by realistic string compactifications
which stabilize all geometrical moduli while simultaneously
addressing the vacuum energy problem of supersymmetry breaking. As
such they are important cases for phenomenological investigation in
their own right and have justly received much attention. More
generally, if we are indeed on the eve of a new supersymmetric data
era then it is of the utmost importance to begin the study of how
this wealth of data can be used to understand the underlying
principles of supersymmetry breaking. The deflected mirage paradigm
provides a framework that is remarkably rich, allowing a smooth
interpolation between supergravity (modulus) mediation, anomaly
mediation, and gauge mediation. It is an important empirical
question to ask how well we can isolate these various contributions
to the transmission of supersymmetry breaking to the observable
sector if all three are present at roughly the same mass scale.

Ultimately we would like to answer this question in as
model-independent a manner as possible, a point emphasized by Choi
and Nilles~\cite{Choi:2007ka}. As those authors point out, gaugino
masses are a much cleaner window on the nature of supersymmetry
breaking and transmission than the rest of the soft
supersymmetry-breaking Lagrangian. This remains true in the
deflected mirage paradigm as well. For this reason we have chosen to
begin answering the question of the previous paragraph with dark
matter observations which depend strongly on the gaugino sector and
relatively less strongly on the scalar sector. Our results indicate
that loosening the restrictions imposed by the original mirage
pattern allows for an expanded parameter space in which the LSP can
be predominantly wino-like, or mixed Higgsino/gaugino. This regime
roughly corresponds to circumstances in which $0 \lappeq N_m
\,\times\, \alpha_g \lappeq 10$, or where $N_m\Lambda_{\rm mess} =
cm_{3/2}$ with $c = \order(1)$. In such cases a number of direct and
indirect detection experiments running now or scheduled to begin
data taking in the near future have excellent prospects for
detecting the relic neutralino. This is also the area in which a
wino-like LSP with roughly the right properties to explain the
PAMELA positron data can be found. For this statement to be borne
out, however, it is still necessary to make modifications to the
standard diffusion parameters and to imagine non-thermal production
mechanisms to explain the present number density. It would be
worthwhile to further investigate the utility of DMM-like models at
fitting the PAMELA positron data.  If the data does indeed seem to
point toward a wino-like LSP, it would be interesting to see how a
mixed-mediation wino arising from the DMM scenario would fare when a
more sophisticated analysis is performed.

Though the predictions from the deflected mirage mediation model
depend strongly on the presence and magnitude of the gauge-mediation
sector, it is not necessarily the case that observations in the dark
matter arena are in and of themselves sufficiently powerful to
determine this fact. That dark matter observations should suffer
from such an ``inverse problem'' is perhaps not surprising. In fact,
the gaugino sector of the DMM scenario is still a two-parameter
family of models and is thus not fully general. Any trio of gaugino
masses in the DMM theory space can be mapped to an equivalent trio
on an un-deflected model. Thus dark matter signals will distinguish
between the two paradigms only via the scalar sector, which serves
to determine the LSP properties only via the electroweak
symmetry-breaking conditions. Using this fact to detect the presence
of the gauge messenger sector will prove difficult, but some handle
can be obtained via gamma ray signals. To make optimal use of this
data we would prefer to have some knowledge of the relevant mass
scales for the gaugino sector. For this it is crucial that the
collider signatures for the deflected mirage mediation model be
computed as a function of the contribution of the three mediation
mechanisms.

\section*{Acknowledgements}
We would like to thank Daniel Feldman for helpful advice on certain
technical issues. This work is supported by the National Science
Foundation under the grant PHY-0653587.

\end{document}